\documentclass[pra,twocolumn,showpacs,preprintnumbers,superscriptaddress]{revtex4-2}
\usepackage{xcolor}
\usepackage{times}
\usepackage{bm}
\usepackage{float}
\usepackage{graphicx}
\usepackage{amsbsy}
\usepackage{amsmath}
\usepackage{amsfonts}
\usepackage{amsthm}
\usepackage{placeins}
\usepackage{autobreak}

\begin{document}
	
	\allowdisplaybreaks
	
	\theoremstyle{plain}
	\newtheorem{theorem}{Theorem}
	\newtheorem{lemma}[theorem]{Lemma}
	\newtheorem{corollary}[theorem]{Corollary}
	\newtheorem{proposition}[theorem]{Proposition}
	\newtheorem{conjecture}[theorem]{Conjecture}
	
	\theoremstyle{definition}
	\newtheorem{definition}[theorem]{Definition}
	
	\theoremstyle{remark}
	\newtheorem*{remark}{Remark}
	\newtheorem{example}{Example}
	\title{Detection of the genuine non-locality of any three-qubit state}
	\author{Anuma Garg, Satyabrata Adhikari}
	\email{anumagarg\_phd2k18@dtu.ac.in, satyabrata@dtu.ac.in} \affiliation{Delhi Technological University, Delhi-110042, Delhi, India}
	
	\begin{abstract}
		\noindent It is known that the violation of Svetlichny inequality by any three-qubit state described by the density operator $\rho_{ABC}$ witness the genuine non-locality of $\rho_{ABC}$. But it is not an easy task as the problem of showing the genuine non-locality of any three-qubit state reduces to the problem of a complicated optimization problem. Thus, the detection of genuine non-locality of any three-qubit state may be considered a challenging task. Therefore, we have taken a different approach and derived the lower and upper bound of the expectation value of the Svetlichny operator with respect to any three-qubit state to study this problem. The expression of the obtained bounds depends on whether the reduced two-qubit entangled state is detected by the CHSH witness operator or not. It may be expressed in terms of the following quantities such as (i) the eigenvalues of the product of the given three-qubit state and the composite system of single qubit maximally mixed state and reduced two-qubit state and (ii) the non-locality of reduced two-qubit state. We then achieve the inequality whose violation may detect the genuine non-locality of any three-qubit state. A few examples are cited to support our obtained results. Lastly, we discuss its possible implementation in the laboratory.
	\end{abstract}
	\maketitle
	
	\section{introduction}
	\noindent The correlation statistics between the subsystems obtained after performing a local measurement on the entangled system \cite{nielsen,wilde} may be incompatible with the principle of local realism.   Since Bell's inequality \cite{bell} has been derived using the principle of local realism so the generated correlation may violate Bell's inequality. This type of correlation may be called a non-local correlation \cite{brunner,genovese,popescu}. The generalized form of Bell's inequality that may be realizable in an experiment was given by Clauser et.al. \cite{chsh} and it is popularly known as Bell-CHSH inequality. Freedman and Clauser also have provided strong experimental evidence, using a generalized form of Bell's inequality, against the existence of local hidden-variable theories \cite{clauser}. B. S. Cirelson \cite{cirelson} proved that quantum mechanics allow up to $2\sqrt{2}$ as an upper bound of generalized Bell's inequality. The upper bound of $2\sqrt{2}$ has been achieved by the two-qubit maximally entangled state. In 1982, A. Aspect et.al. \cite{aspect} showed that maximum violation of generalized Bell's inequality can be achieved in an experiment. Later, Horodecki et. al. \cite{horo3} also studied the problem of non-locality for two-qubit states and provided a criterion to check the non-locality of $\rho_{AB}$ in terms of $M(\rho_{AB})$, where  $M(\rho_{AB})$ is the sum of the two largest eigenvalues of $T^{t}T$. $T$ denote the correlation matrix of $\rho_{AB}$. The criterion states that any two-qubit state violates Bell's inequality if and only if M($\rho_{AB})>1$.\\
A lot of research had already been done in studying the problem of two-qubit non-locality \cite{weighs,methot,hoban,collins,liang,kwait,andreoli,pal,acin,batle}. Therefore, researcher turn on to the study of non-locality of  multi-partite state \cite{bancal,mao,curchod,zhang,bancal1,niset}. As the number of qubits increases in the system, the complexity of the system also increases. Therefore, the study of the non-locality of the multipartite system is a difficult problem but in spite of that some progress has been achieved. In particular, the non-locality of the three-qubit system is relatively easier to handle. Non-locality of a three-qubit state can be tested by various inequality such as Svetlichny inequality \cite{svetlichny}, Mermin inequality \cite{mermin} and logical inequality based on GHZ type event probabilities \cite{ren}. The experimental verification of the non-locality of the three-qubit GHZ state is reported in \cite{pan1}. The non-locality of three-qubit pure symmetric state have been explored in \cite{anjali}. The standard non-locality and genuine non-locality of GHZ symmetric state have been studied in \cite{paul}.\\
	Mermin inequality \cite{mermin} can be considered as a generalized form of the Bell-CHSH inequality and it can be violated by not only genuine entangled three-qubit states but also
	by biseparable states. Thus, the discrimination of the classes of three-qubit entangled state is not possible by merely observing the violation of Mermin inequality. But fortunately, there exists another inequality known as Svetlchny inequality \cite{svetlichny} violation which guarantees the fact that the three-qubit state under investigation is a genuine entangled state. Therefore, the genuine tripartite nonlocal correlation that may exist in the three-qubit state $\rho_{ABC}$ may be detected by Svetlichny inequality, which is given by \cite{svetlichny}
	\begin{eqnarray}
		|\langle S_{v}\rangle_{\rho_{ABC}}|\leq 4
		\label{svetineq}
	\end{eqnarray}
	where $S_{v}$ denote the Svetlichny operator, which may be defined as
	\begin{eqnarray}
		S_{v}&=&\vec{a}.\vec{\sigma_{1}}\otimes [\vec{b}.\vec{\sigma_{2}}\otimes (\vec{c}+\vec{c'}).\vec{\sigma_{3}}+\vec{b'}.\vec{\sigma_{2}}\otimes (\vec{c}-\vec{c'}).\vec{\sigma_{3}}]\nonumber\\&+& \vec{a'}.\vec{\sigma_{1}}\otimes [\vec{b}.\vec{\sigma_{2}}\otimes (\vec{c}-\vec{c'}).\vec{\sigma_{3}}-\vec{b'}.\vec{\sigma_{2}}\otimes (\vec{c}+\vec{c'}).\vec{\sigma_{3}}]\nonumber \\
	\end{eqnarray}
	Here $\vec{a},\vec{a'}$; $\vec{b},\vec{b'}$ and
	$\vec{c},\vec{c'}$ are the unit vectors and the $\vec{\sigma_{i}}=(\sigma_{i}^{x},\sigma_{i}^{y},\sigma_{i}^{z})$ denote the spin projection operators. To obtain the maximal violation of the Svetlichny inequality, the expectation value of the Svetlichny operator must achieve the value $4\sqrt{2}$. In particular, the violation of Svetlichny inequality by three-qubit generalized GHZ state,  maximal slice state, and W class state has been studied in \cite{ghose,rungta1} and it has been found that the maximal violation $4\sqrt{2}$ may be obtained for $GHZ$ state. The theoretical result of Ghose et.al. has been demonstrated experimentally in \cite{lu}. An operational method to detect the genuine multipartite non-locality for three-qubit mixed states has been investigated in \cite{mli}. Also, the genuine nonlocality of three-qubit pure and mixed states has been extensively studied in \cite{lysun}. \\
	In order to obtain the violation of the Svetlichny inequality, one has to calculate the expectation of the Svetlichny operator by maximizing overall measurements of spin in the directions $\vec{a},\vec{a'}, \vec{b},\vec{b'},\vec{c},\vec{c'}$. Consequently, the problem of the violation of the Svetlichny inequality reduces to an optimization problem, which is not very easy to solve for any arbitrary three-qubit state. This motivates us to find a way by which we can overcome this problem. To achieve our task, we derive the upper and lower bound of the expectation value of the Svetlichny operator with respect to any three-qubit state. These newly obtained upper and lower bounds depends on the non-locality of the reduced two-qubit state of the three-qubit system and we have shown that this may pave the way to study the genuine non-locality of any three-qubit state.\\
	This work can be organized as follows: In section-II, we have provided a short summary of results and concepts developed in earlier works. In section-III, the lower and upper bound of the expectation value of the Svetlichny operator is obtained, which may be considered the main ingredient to fulfill the motivation of this work. In section-IV, we have derived the inequality whose violation guarantees the genuine non-locality of any three-qubit state. In section-V, we have illustrated our result with a few examples. Lastly, we have provided the conclusion. \\   
	
	\section{Recapitulation}
	\noindent In this section, we have re-stated an important inequality and a corollary and then revisited the non-locality of the two-qubit state studied in \cite{anuma}. These ingredients may be used as a tool to develop the content of the later section.
	\subsection{A few Results}
	\noindent \textbf{R-1:}
	For M be any $n\times n$ complex matrix and N be any $n\times n$ Hermitian matrix, we have \cite{horn,lasserre}
	\begin{eqnarray}
		\lambda_{min}(\overline{M})Tr(N)\leq R(Tr(MN))\leq \lambda_{max}(\overline{M})Tr(N)
		\label{r1}
	\end{eqnarray}
where $\overline{M}=\frac{M+M^{\dagger}}{2}$ and R(x) denotes the real part of x.\\
\textbf{Proof:-} Let us assume that the eigenvalues of $\overline{M}$ may be arranged in an ascending order as $\lambda_{min}= \lambda_{1} \leq \lambda_{2} \leq ......\leq \lambda_{n}=\lambda_{max}$. To prove R-1, let us recall the lower and upper bound of R(Tr(MN)) which is given in \cite{lasserre}, 
\begin{eqnarray}
	\sum_{i=1}^{n}\lambda_{i}(\overline{M}) \lambda_{n-i+1}(N) \leq R(Tr(MN))\leq \sum_{i=1}^{n}\lambda_{i}(\overline{M}) \lambda_{i}(N)	
\end{eqnarray}
 In L.H.S., Replacing all the eigenvalues of $\overline{M}$ by its minimum eigen values and in R.H.S. if we replace all the eigenvalues of $\overline{M}$ by its maximum eigenvalue. We get the desired result given in (\ref{r1}).\\

\noindent \textbf{Cor-1:}
For M be any $n\times n$ complex matrix and N be any $n\times n$ Hermitian matrix, we have 
\begin{eqnarray}
	Tr(\overline{M})\lambda_{min}(N)\leq R(Tr(MN))\leq Tr(\overline{M})\lambda_{max}(N)
	\label{r2}
\end{eqnarray}
where $\overline{M}=\frac{M+M^{\dagger}}{2}$ and R(x) denotes the real part of x.\\

	
		\noindent \textbf{Cor-2:}
	For M be any $n\times n$ complex matrix and N be any $n\times n$ Hermitian matrix, we have
	\begin{eqnarray}
		Tr(\overline{M})\lambda_{k}(N)\leq R(Tr(MN))
		\label{cor2}
	\end{eqnarray}
	where $\lambda_{k}(N)$ denote the first non-zero eigenvalue of $N$.
	\subsection{Revisiting the nonlocality of two-qubit entangled states determined by $S_{NL}$}
	\noindent Let us consider an XOR Game played between two distinct players Alice(A) and Bob(B) and suppose that they share a two-qubit state $\rho_{AB}$ which is given in the form \cite{luo}
	\begin{equation}
		\rho_{AB}= \frac{1}{4}[I\otimes I+ \overrightarrow{a}.\overrightarrow{\sigma}\otimes I +I \otimes \overrightarrow{b}.\overrightarrow{\sigma}+ \sum c_{j} \sigma_{j} \otimes \sigma_{j}]
		\label{gen2qbitst22}
	\end{equation} 
	where $c_{i}\in R$ and $\sigma_{i}$ denote the Pauli matrices and the unit vectors $\vec{a}$ and $\vec{b}$ are given by $\vec{a}\equiv (a_{1},a_{2},a_{3})\in R^{3}$ and $\vec{b}\equiv (b_{1},b_{2},b_{3})\in R^{3}$.\\
	If players A and B play the game using the shared state $\rho_{AB}$ then the maximum probability $P^{max}$ of winning the game overall strategy is given by\cite{jon,archan}
	\begin{eqnarray}
		P^{max}&=&\frac{1}{2}[1+\frac{\langle B_{CHSH}\rangle _{\rho_{AB}}}{4}]
		\label{maxprob}
	\end{eqnarray}
	where $B_{CHSH}=A_{0}\otimes B_{0}+A_{0}\otimes B_{1}+A_{1}\otimes B_{0}-A_{1}\otimes B_{1}$ denote the Bell-CHSH operator and $A_{0}$, $B_{0}$, $A_{1}$ and $B_{1}$ denote the Hermitian operators. Also we have $\langle B_{CHSH}\rangle _{\rho_{AB}}=Tr[(A_{0}\otimes B_{0}+A_{0}\otimes B_{1}+A_{1}\otimes B_{0}-A_{1}\otimes B_{1})\rho_{AB}]$.\\
	The strength of the non-locality of $\rho_{AB}$ denoted by $S_{NL}(\rho_{AB})$ in terms of $P^{max}$ may be defined as 
	\begin{equation}
		S_{NL}(\rho_{AB})=max \{P^{max}-\frac{3}{4},0\}
		\label{snl22}
	\end{equation}
	For any classical theory, $P^{max}\leq \frac{3}{4}$ and hence $S_{NL}(\rho_{AB})=0$. For quantum mechanical theory and for non-signaling correlation, we have $P^{max} > \frac{3}{4}$   and thus $S_{NL}(\rho_{AB})\neq 0$.
	We have further considered different measurement setting $xy-$, $xz-$, and $yz-$ planes to calculate the maximum probability of success of winning the game. In these planes, the maximum probability of success of winning the game is denoted by $P_{xy}$, $P_{xz}$, and $P_{yz}$ respectively. Therefore, the corresponding maximum probability of success may be expressed as
	\begin{equation}
		P_{ij}^{max}=\frac{1}{2}[1+\frac{\langle B^{(ij)}_{CHSH}\rangle_{\rho_{AB}}}{4}] ,i,j=x,y,z~~ \&~~  i\neq j
		\label{R1}
	\end{equation}   
	The Bell operators $B^{(xy)}_{CHSH}$, $B^{(xz)}_{CHSH}$ and $B^{(yz)}_{CHSH}$ can be written in terms of the observables $\sigma_{x}$, $\sigma_{y}$, and $\sigma_{z}$ as\cite{hyllus} 
	\begin{eqnarray}
		B^{(ij)}_{CHSH}&=&\sigma_{i}\otimes \frac{\sigma_{i}+\sigma_{j}}{\sqrt{2}}+ \sigma_{i}\otimes \frac{\sigma_{i}-\sigma_{j}}{\sqrt{2}}\nonumber\\&+&
		\sigma_{j}\otimes \frac{\sigma_{i}+\sigma_{j}}{\sqrt{2}}- \sigma_{j}\otimes \frac{\sigma_{i}-\sigma_{j}}{\sqrt{2}},\nonumber\\&& i,j=x,y,z~~ \&~~ i\neq j
		\label{bchsh}
	\end{eqnarray}
	In terms of $P_{xy}$, $P_{xz}$ and $P_{yz}$, the strength of the non-locality $S_{NL}^{(ij)}(\rho_{AB})~(i,j=x,y,z,~~i\neq j)$, may be defined as
	\begin{eqnarray}
		S_{NL}^{(ij)}(\rho_{AB})=max\{P,0\}, i,j=x,y,z,~~i\neq j
		\label{snl23}
	\end{eqnarray}
	where $P=\{P^{max}_{xy}-\frac{3}{4}, P^{max}_{xz}-\frac{3}{4}, P^{max}_{yz}-\frac{3}{4}\}$.\\ 
	The strength of the non-locality $S_{NL}(\rho_{AB})$ can also be expressed in terms of the CHSH witness operator and therefore, the expression of $S_{NL}(\rho_{AB})$ may be given in the following result.\\
	\textbf{Result-1:-} If $\rho_{AB}$ denote any arbitrary two-qubit bipartite state shared between the two distant players Alice and Bob and if $B_{CHSH}$ represent the CHSH operator and $W_{CHSH}=2I-B_{CHSH}$ denote the CHSH witness operator then the strength of the non-locality $S_{NL}(\rho_{AB})$ may be given by 
	\begin{eqnarray}
		S_{NL}(\rho_{AB})=max \{-\frac{Tr[W_{CHSH}\rho_{AB}]}{8},0\}
		\label{snl1}		
	\end{eqnarray}
	Further, if the game is played with a shared two-qubit state $\rho_{AB}$ and if measurement is performed in different planes such as $xy-$, $yz-$ and $zx-$ plane then 
	\begin{equation}
		P^{max}_{ij}=\frac{3}{4}-\frac{Tr[W_{CHSH}^{(ij)}\rho_{AB}]}{8} ,i,j=x,y,z~~ \&~~  i\neq j
		\label{R2}
	\end{equation} 
	Therefore, the strength of the nonlocality $S_{NL}^{(ij)}(\rho_{AB})$  may be re-defined in terms of witness operator $W_{CHSH}^{(ij)}$ as
	\begin{eqnarray}
		S_{NL}^{(ij)}(\rho_{AB})=max \{-\frac{Tr[W^{ij}_{CHSH}\rho_{AB}]}{8},0\}
		\label{snl2}		
	\end{eqnarray}
	We should note here a few important facts regarding the strength of the non-locality of $\rho_{AB}$, which is given below:\\ 
	(F1) If $P^{max}>\frac{3}{4}$ then $S_{NL}(\rho_{AB})\neq 0$.\\
	(F2) If the players are playing the game with a separable state $\rho_{AB}^{(sep)}$ then $P^{max}\in [0,\frac{3}{4}]$ and thus $S_{NL}(\rho_{AB})=0$.\\
	(F3) It is known that there exists two-qubit entangled states $\rho_{AB}^{(ent)}$ which does not violate the CHSH inequality and thus from (\ref{maxprob}), we have $P^{max}\leq\frac{3}{4}$. Therefore, $\rho_{AB}^{(ent)}$ is not detected by the CHSH witness operator $W_{CHSH}$ and hence $Tr[W_{CHSH}\rho_{AB}^{(ent)}]\geq 0$. We may now conclude from the definition (\ref{snl1}) of the strength of the non-locality that $S_{NL}(\rho_{AB}^{ent})=0$.
	By going through the above facts, it may be easily seen that the conclusion made in (F3) is not correct since the state under consideration in (F3) is an entangled state.  Therefore, we may infer that the definition of the strength of the non-locality given in (\ref{snl1}) does not hold good when CHSH witness operator $W_{CHSH}$ does not detect the entangled state $\rho_{AB}^{(ent)}$. Thus, to resolve this problem, we consider an entangled state $\rho_{AB}^{ent}$, which is not detected by $W_{CHSH}$ and re-define the strength of its nonlocality as 
	\begin{eqnarray}
		S^{New}_{NL}(\rho_{AB}^{ent})&=&r(P^{max}-\frac{3}{4})+(1-r)K
		\label{snldef}
	\end{eqnarray}
	where $ 0\leq r<1$ and $K=\frac{Tr[W_{CHSH}\rho_{AB}(\rho_{AB})^{T_{B}}]}{4N(\rho_{AB})}$.\\
	The parameter $r$ is chosen in such a way that $S^{New}_{NL}(\rho_{AB}^{ent})>0$. The upper bound of $r$ can be obtained by imposing the restriction $S^{New}_{NL}(\rho_{AB})>0$ and it is given by
	\begin{eqnarray}
		r<\frac{K}{\frac{3}{4}-P^{max}+K}
		\label{qub1}
	\end{eqnarray}
	Therefore, in this way, we have shown in \cite{anuma} that it is possible to calculate the strength of non-locality of any bipartite two-qubit entangled state whether it is detected/not detected by the CHSH witness operator.\\
	Furthermore, we have derived the relation between the non-locality of pure three-qubit entangled state in terms of the strength of non-locality of two-qubit entangled state. In \cite{anuma}, the relations are derived for a few three-qubit pure entangled states. In this work, we will generalize the relationship between the non-locality of any arbitrary three-qubit state and the non-locality of its reduced two-qubit state.
	\section{Lower and Upper bound of the expectation value of the Svetlichney operator}
	\noindent In this section, we construct the Hermitian operators to derive a connection between the two-qubit nonlocality determined by the strength of the nonlocality $S_{NL}$ and the nonlocality of an arbitrary (either pure or mixed) three-qubit state determined by the Svetlichney operator $S_{v}$. The construction of the Hermitian operator makes us enable to derive the lower and upper bound of the expectation value of the Svetlichney operator with respect to an arbitrary three-qubit state. The derived bound of the expectation value of the Svetlichney operator provides us a new way to discriminate the genuine three-qubit entangled state. \\ 
	To proceed forward, let us consider a three-qubit state (pure or mixed) described by the density operator $\rho_{ABC}$ and its reduced two-qubit entangled state $\rho_{ij},~i,j=A,B,C~ and~ i\neq j$, which can be related by the following way:
	\begin{eqnarray}
		\rho_{ij}=Tr_{k}[\rho_{ABC}],~i,j,k=A,B,C~and~i\neq j \neq k 
		\label{forma}
	\end{eqnarray}
	The two operators may be constructed as 
	\begin{eqnarray}
		A_{l}=pS_{v}+(1-p)(I_{2}\otimes W_{CHSH})
		\label{cons1a}
	\end{eqnarray}
	\begin{eqnarray}
		B_{l}=\rho_{ABC}(I_{2}\otimes \rho_{ij}),~i,j=A,B,C,~i\neq j
		\label{cons1b}
	\end{eqnarray}
	where $p \in [0,1]$ and $W_{CHSH}(=2I_{2}-B_{CHSH})$ denote the CHSH witness operator. $I_{2}$ denotes the identity matrix of order 2. Now, in the subsequent subsections, we derive the lower and upper bound of the expectation value of the Svetlichney operator in terms of two-qubit non-locality determined by $S_{NL}(\rho_{ij})$.\\ 
	
	\subsection{Lower bound of the expectation value of Svetlichney operator in terms of two-qubit non-locality determined by $S_{NL}$}
	\noindent To derive the lower bound of the expectation value of Svetlichney operator $S_{v}$, let us start with the quantity $R(Tr[A_{l}B_{l}])$. It can be expressed as
	\begin{eqnarray}
		R(Tr[A_{l}B_{l}])&=&R(Tr[(pS_{v}+(1-p)(I_{2}\otimes W_{CHSH}))\times \nonumber\\&&\rho_{ABC}(I_{2}\otimes \rho_{ij})])\nonumber \\  
		&=& pR(Tr[S_{v}\rho_{ABC}(I_{2}\otimes \rho_{ij})])+(1-p)\times \nonumber\\ &&R(Tr[(I_{2}\otimes\rho_{ij} W_{CHSH})\rho_{ABC}])
		\label{trab}
	\end{eqnarray}
	Since $(I_{2}\otimes \rho_{ij})$ and $\rho_{ABC}$ is a hermitian operator, and $S_{v}\rho_{ABC}$ and $(I_{2}\otimes\rho_{ij} W_{CHSH})$ are complex matrices so after applying \textbf{Cor-1} on (\ref{trab}), we get
	\begin{eqnarray}
		R(Tr[S_{v}\rho_{ABC}(I_{2}\otimes \rho_{ij})])\leq \lambda_{max}(I_{2}\otimes \rho_{ij})Tr[\overline{S_{v}\rho_{ABC}}]\nonumber \\
		\label{her1}
	\end{eqnarray}
	\begin{eqnarray}
		R(Tr[(I_{2}\otimes\rho_{ij} W_{CHSH})\rho_{ABC}]) &\leq& Tr[\overline{I_{2}\otimes\rho_{ij} W_{CHSH}}]\times \nonumber\\&&\lambda_{max}(\rho_{ABC})
		\label{her2}
	\end{eqnarray}
	Using (\ref{her1}) and (\ref{her2}) in (\ref{trab}), we obtain
	\begin{eqnarray}
		R(Tr[A_{l}B_{l}])&=& R(pTr[S_{v}\rho_{ABC}(I_{2}\otimes \rho_{ij})]) +(1-p)\times \nonumber\\&&R(Tr[(I_{2}\otimes\rho_{ij} W_{CHSH})\rho_{ABC}])\nonumber\\ &\leq& p\lambda_{max}(I_{2}\otimes \rho_{ij})Tr[\overline{S_{v}\rho_{ABC}}] \nonumber \\ &+& (1-p)\lambda_{max}(\rho_{ABC})Tr[\overline{I_{2}\otimes\rho_{ij} W_{CHSH}}] \nonumber\\&=& p\lambda_{max}(I_{2}\otimes \rho_{ij})\langle S_{v}\rangle_{\rho_{ABC}} \nonumber\\ &+&2(1-p)Tr[ W_{CHSH}\rho_{ij}]\lambda_{max}(\rho_{ABC})
		\label{ineq200}
	\end{eqnarray}
	In the last step, one can easily check that $Tr[\overline{S_{v}\rho_{ABC}}]$= $Tr[S_{v}\rho_{ABC}]$, $Tr[\overline{I_{2}\otimes\rho_{ij} W_{CHSH}}]= Tr[I_{2}\otimes\rho_{ij} W_{CHSH}]$, and $Tr[I_{2}\otimes\rho_{ij} W_{CHSH}]=2Tr[W_{CHSH}\rho_{ij}]$.
	\\
	Again applying LHS of \textbf{R-1} on Hermitian operator $pS_{v}+(1-p)(I_{2}\otimes W_{CHSH}))$, and $\rho_{ABC}(I_{2}\otimes \rho_{ij})$ be any complex matrix and using $Tr[S_{v}]=0$, we get
	\begin{eqnarray}
		&&R(Tr[(pS_{v}+(1-p)(I_{2}\otimes W_{CHSH}))\rho_{ABC}(I_{2}\otimes \rho_{ij})])\nonumber \\  
		&&\geq Tr[pS_{v}+(1-p)(I_{2}\otimes W_{CHSH})] \lambda_{min}(\overline{\rho_{ABC}(I_{2}\otimes \rho_{ij})})\nonumber  \\
		&&= 8(1-p)\lambda_{min}(\overline{\rho_{ABC}(I_{2}\otimes \rho_{ij})})
		\label{ineq201}
	\end{eqnarray}
	In the second line of (\ref{ineq201}), we have used the linearity property of trace and $Tr( W_{CHSH})=4$, where $W_{CHSH}=2I-B_{CHSH}$. Combining the inequalities (\ref{ineq200}) and (\ref{ineq201}), we get  
	\begin{eqnarray}
		&& [p\lambda_{max}(I_{2}\otimes \rho_{ij})\langle S_{v}\rangle_{\rho_{ABC}}+2(1-p)Tr[ W_{CHSH}\rho_{ij}]\times \nonumber \\ && \lambda_{max}(\rho_{ABC})]
		\geq 8(1-p)\lambda_{min}(\overline{\rho_{ABC}(I_{2}\otimes \rho_{ij})})
		\label{ineq203}
	\end{eqnarray}
	After simplification, the inequality (\ref{ineq203}) can be re-expressed as
	\begin{eqnarray}
		\langle S_{v}\rangle_{\rho_{ABC}}&\geq& \frac{8(1-p)\lambda_{min}(\overline{\rho_{ABC}(I_{2}\otimes \rho_{ij})})}{p\lambda_{max}(I_{2}\otimes \rho_{ij})}\nonumber \\&-&\frac{2(1-p)Tr[ W_{CHSH}\rho_{ij}]\lambda_{max}(\rho_{ABC})}{p\lambda_{max}(I_{2}\otimes \rho_{ij})} 
		\label{sv}
	\end{eqnarray}
	Since our aim is to establish the relationship between $\langle S_{v}\rangle_{\rho_{ABC}}$ and the strength of the nonlocality $S_{NL}(\rho_{ij})$ of two-qubit entangled state $\rho_{ij}$ so we shall consider two cases in which we discuss the following: (i) When $\rho_{ij}$ is detected by the witness operator $W_{CHSH}$ and (ii) When $W_{CHSH}$ does not detect the state $\rho_{ij}$.   
	\subsubsection{When the entangled state $\rho_{ij}$ is detected by the witness operator $W_{CHSH}$}
	\noindent Let us recall the definition (\ref{snl1}) of $S_{NL}(\rho_{ij})$ and can be re-expressed for the entangled state $\rho_{ij}$ as  
	\begin{eqnarray}
		S_{NL}(\rho_{ij})=\frac{-Tr[W_{CHSH}\rho_{ij}]}{8} \nonumber
		\label{snlrho12}
	\end{eqnarray}
	Putting this value of $S_{NL}(\rho_{ij})$ in (\ref{sv}), we get
	\begin{eqnarray}
		\langle S_{v}\rangle_{\rho_{ABC}}&\geq& \frac{8(1-p)\lambda_{min}(\overline{\rho_{ABC}(I_{2}\otimes \rho_{ij})})}{p\lambda_{max}(I_{2}\otimes \rho_{ij})}\nonumber \\ 
		&+&\frac{16(1-p)\lambda_{max}(\rho_{ABC})S_{NL}(\rho_{ij})}{p\lambda_{max}(I_{2}\otimes \rho_{ij})}
		\label{lb1} 
	\end{eqnarray}
	\subsubsection{When $W_{CHSH}$ does not detected the entangled state $\rho_{ij}$}
	\noindent In this case, $S_{NL}(\rho_{ij})$ is defined in a different way and it is given by (\ref{snldef})
	\begin{eqnarray}
		S^{New}_{NL}(\rho_{ij})=r(P^{max}-\frac{3}{4})+(1-r)K
		\label{snlcase2}
	\end{eqnarray}
	where $K$ is defined as 
	\begin{eqnarray}
		K=\frac{Tr[W_{CHSH}\rho_{ij}(\rho_{ij}^{T_{j}})]}{4N(\rho_{ij})}
		\label{k1}
	\end{eqnarray}
	where $T_{j}$ represent the partial transposition with respect to the qubit "j" and $N(\rho_{ij})$ denote the negativity of the two-qubit entangled state $\rho_{ij}$.\\
	To derive the lower bound of $\langle S_{v}\rangle$ for this case, we need a lemma which can be stated as:\\ 
	\textbf{Lemma-1:} If an entangled state described by the density operator $\rho_{ij}$ and the witness operator $W_{CHSH}$ does not detect it then 
	\begin{eqnarray}
		K\geq\frac{\lambda_{min}[(\rho_{ij}^{T_{j}})^{2}]Tr[W_{CHSH}\rho_{ij}]}{4\lambda_{max}[\rho_{ij}^{T_{j}}]N(\rho_{ij})}
		\label{lemma}
	\end{eqnarray}
	where $K$ is given by $(\ref{k1})$.\\
	The proof of the $\textbf{Lemma-1}$ can be found in the Appendix-I.\\
	Now we are in a position to establish the relationship between  $S^{New}_{NL}(\rho_{ij})$ and $\langle S_{v}\rangle_{\rho_{ABC}}$ when the witness operator $W_{CHSH}$ does not detect the entangled state $\rho_{ij}$.\\
	Using (\ref{lemma}), the expression for the strength of the nonlocality $S^{New}_{NL}(\rho_{ij})$ given in (\ref{snlcase2}) can be written as
	\begin{eqnarray}
		S^{New}_{NL}(\rho_{ij})&\geq& r(P^{max}-\frac{3}{4})\nonumber \\ &+&(1-r)\frac{Tr[W_{CHSH}\rho_{ij}]\lambda_{min}[(\rho_{ij}^{T_{j}})^{2}]}{4\lambda_{max}[\rho_{ij}^{T_{j}}]N(\rho_{ij})}
		\label{600}
	\end{eqnarray}
	The above inequality (\ref{600}) may be re-expressed as 
	\begin{eqnarray}
		Tr[W_{CHSH}\rho_{ij}]&\leq&\frac{4Z[S^{New}_{NL}(\rho_{ij})-r(P^{max}-\frac{3}{4})]}{(1-r)\lambda_{min}[(\rho_{ij}^{T_{j}})^{2}]}
		\label{ineqwchsh}
	\end{eqnarray}
	where $Z=N(\rho_{ij})\lambda_{max}[\rho_{ij}^{T_{j}}]$.\\
	Using the inequality (\ref{ineqwchsh}) in (\ref{sv}), we get
	\begin{eqnarray}
		\langle S_{v}\rangle_{\rho_{ABC}}&\geq& 8(1-p)[\frac{\lambda_{min}(\overline{\rho_{ABC}(I_{2}\otimes \rho_{ij}))}}{p\lambda_{max}(I_{2}\otimes \rho_{ij})}\nonumber\\&-&G]  
	\end{eqnarray} 
	where $G=\frac{\lambda_{max}(\rho_{ABC})[S^{New}_{NL}(\rho_{ij})-r(P^{max}-\frac{3}{4})]N(\rho_{ij})\lambda_{max}[\rho_{ij}^{T_{j}}]}{p(1-r)\lambda_{min}[(\rho_{ij}^{T_{j}})^{2}]\lambda_{max}(I_{2}\otimes \rho_{ij})}$.\\
	We are now in a position to collect all the above obtained results in the following theorem:\\
	
\noindent	\textbf{Theorem-1a:} The lower bound of the expectation value of the Svetlichny operator $S_{v}$ with respect to three-qubit state $\rho_{ABC}$ is given by\\
	\begin{eqnarray}
		(i)~~\langle S_{v}\rangle_{\rho_{ABC}} &\geq& \frac{8(1-p)\lambda_{min}(\overline{\rho_{ABC}(I_{2}\otimes \rho_{ij})})}{p\lambda_{max}(I_{2}\otimes \rho_{ij})}\nonumber\\&+&\frac{16(1-p)\lambda_{max}(\rho_{ABC})S_{NL}(\rho_{ij})}{p\lambda_{max}(I_{2}\otimes \rho_{ij})} 
		\label{svlb1}
	\end{eqnarray}
	and
	\begin{eqnarray}
		(ii)~~\langle S_{v}\rangle_{\rho_{ABC}}&\geq& 8(1-p)\biggl[\frac{\lambda_{min}(\overline{\rho_{ABC}(I_{2}\otimes \rho_{ij})})}{p\lambda_{max}(I_{2}\otimes \rho_{ij})}-\nonumber\\&&\frac{\big(S^{New}_{NL}(\rho_{ij})-r(P^{max}-\frac{3}{4})\big)\times A_{1}}{p(1-r)\lambda_{min}[(\rho_{ij}^{T_{j}})^{2}]\lambda_{max}(I_{2}\otimes \rho_{ij})}\biggr] \nonumber\\
		\label{svlb2}
	\end{eqnarray}
where $A_{1}=(N(\rho_{ij})\lambda_{max}(\rho_{ij}^{T_{j}})\lambda_{max}(\rho_{ABC}))$\\
	according as when the entangled state $\rho_{ij}$ does or does not detected by the witness operator $W_{CHSH}$.
	\subsection{Upper bound of the expectation value of Svetlichney operator in terms of two-qubit non-locality determined by $S_{NL}$}
	\noindent Let us consider two operators $A_{u}$ and $B_{u}$ which may be defined as 
	\begin{eqnarray}
		A_{u}=qS_{v}+(1-q)(I_{2}\otimes W_{CHSH}),~~0\leq q \leq 1
		\label{cons2a}
	\end{eqnarray}  
	\begin{eqnarray}
		B_{u}=\rho_{ABC}(I_{2}\otimes \rho_{ij})
		\label{cons2b}
	\end{eqnarray}
	The expression for $R(Tr[A_{u}B_{u}])$ is given by
	\begin{eqnarray}
		&&R(Tr[(qS_{v}+(1-q)(I_{2}\otimes W_{CHSH}))\rho_{ABC}(I_{2}\otimes \rho_{ij})])\nonumber\\  
		&=&qR(Tr[S_{v}\rho_{ABC}(I_{2}\otimes \rho_{ij})]) +(1-q)R(Tr[(I_{2}\otimes W_{CHSH}) \nonumber\\&& \times \rho_{ABC}(I_{2}\otimes \rho_{ij})]) \nonumber  \\
		&=&qR(Tr[S_{v}\rho_{ABC}(I_{2}\otimes \rho_{ij})]) +(1-q)R(Tr[(I_{2}\otimes\rho_{ij} \nonumber \\ &&\times W_{CHSH}) \rho_{ABC}]) \nonumber \\&\geq&
		qR(Tr[S_{v}\rho_{ABC}(I_{2}\otimes \rho_{ij})]) +(1-q)Tr[\overline{(I_{2}\otimes\rho_{ij} W_{CHSH})}]\nonumber \\ &&\times\lambda_{min}(\rho_{ABC})\nonumber \\
		&=& qR(Tr[S_{v}\rho_{ABC}(I_{2}\otimes \rho_{ij})]) +2(1-q)Tr[W_{CHSH}\rho_{ij} ]\times \nonumber \\ &&\lambda_{min}(\rho_{ABC}) 
		\label{7}  
	\end{eqnarray}
	In the second and third lines, we have used the linearity and cyclic property of the trace. We have used the LHS inequality of $\textbf{Cor-1}$ on Hermitian operator $\rho_{ABC}$ and considering ($I_{2}\otimes\rho_{ij} W_{CHSH})$ be any complex matrix in the fourth line. In the last line, we have used $Tr[I_{2}\otimes\rho_{ij} W_{CHSH} ]=Tr[\overline{I_{2}\otimes\rho_{ij} W_{CHSH}} ]$ and one of the properties of the trace i.e. $Tr[(I_{2}\otimes\rho_{ij} W_{CHSH})]$=2$Tr[W_{CHSH}\rho_{ij}]$.\\   
	Applying RHS inequality of $\textbf{R-1}$ on the Hermitian operator $qS_{v}+(1-q)(I_{2}\otimes W_{CHSH}))$ and considering $\rho_{ABC}(I_{2}\otimes \rho_{ij})$ be any complex matrix, we get 
	\begin{eqnarray}
		&&R(Tr[(qS_{v}+(1-q)(I_{2}\otimes W_{CHSH}))\rho_{ABC}(I_{2}\otimes \rho_{ij})])\nonumber \\  
		\leq&&Tr[qS_{v}+(1-q)(I_{2}\otimes W_{CHSH})] \lambda_{max}(\overline{\rho_{ABC}(I_{2}\otimes \rho_{ij})})\nonumber  \\=&& 8(1-q)\lambda_{max}(\overline{\rho_{ABC}(I_{2}\otimes \rho_{ij})})
		\label{8}
	\end{eqnarray}
	In the third line, we find $Tr[S_{v}]=0$ and $Tr[I_{2}\otimes W_{CHSH}]=8$.\\
	Combining (\ref{7}) and (\ref{8}), we get  
	\begin{eqnarray}
		&& qR(Tr[S_{v}\rho_{ABC}(I_{2}\otimes \rho_{ij})]) +2(1-q)Tr[W_{CHSH}\rho_{ij}]\times \nonumber\\&&\lambda_{min}(\rho_{ABC}) \nonumber \\ 
		&\leq& 8(1-q)\lambda_{max}(\overline{\rho_{ABC}(I_{2}\otimes \rho_{ij})})
		\label{9}
	\end{eqnarray}
	Again using $\textbf{Cor-2}$ on Hermitian operators  $I_{2}\otimes \rho_{ij}$ and $S_{v}\rho_{ABC}$ be any complex matrix, we get
	\begin{eqnarray}
		Tr[\overline{S_{v}\rho_{ABC}}]\lambda_{k}((I_{2}\otimes \rho_{ij})) \leq R(Tr[S_{v}\rho_{ABC}(I_{2}\otimes \rho_{ij})]) \nonumber \\ 
		\implies Tr[S_{v}\rho_{ABC}]\lambda_{k}((I_{2}\otimes \rho_{ij})) \leq R(Tr[S_{v}\rho_{ABC}(I_{2}\otimes \rho_{ij})]) \nonumber\\
		\label{r21}
	\end{eqnarray}
	where $Tr[\overline{S_{v}\rho_{ABC}}]= Tr[S_{v}\rho_{ABC}]$.\\
	Using (\ref{r21}), the inequality (\ref{9}) may be re-expressed as
	\begin{eqnarray}
		\langle S_{v}\rangle_{\rho_{ABC}}&\leq& \frac{8(1-q)\lambda_{max}(\overline{\rho_{ABC}(I_{2}\otimes \rho_{ij})})}{q\lambda_{k}(I_{2}\otimes \rho_{ij})}\nonumber \\ 
		&-&\frac{2(1-q)Tr[W_{CHSH}\rho_{ij} ]\lambda_{min}(\rho_{ABC})}{q\lambda_{k}(I_{2}\otimes \rho_{ij})} 
		\label{upperbound}
	\end{eqnarray}
	where $\lambda_{k}(I_{2}\otimes \rho_{ij})$ is the first non-zero eigenvalue of $(I_{2}\otimes \rho_{ij}).$\\ 
	The upper bound (\ref{upperbound}) of the expectation value of the operator $S_{v}$ with respect to any three-qubit state $\rho_{ABC}$ can be further studied in terms of the non-locality $S_{NL}(\rho_{ij})$ of two-qubit state by considering the following two cases: (i) When the state $\rho_{ij}$ is detected by $W_{CHSH}$ and (ii) When the state $\rho_{ij}$ is not detected by $W_{CHSH}$.
	\subsubsection{When the state $\rho_{ij}$ is detected by $W_{CHSH}$ }
	\noindent In this case, we are considering the two-qubit entangled state $\rho_{ij}$, which is detected by the witness operator $W_{CHSH}$. Therefore, using the definition of $S_{NL}(\rho_{ij})$ given in (\ref{snl1}), the inequality (\ref{upperbound}) reduces to
	\begin{eqnarray}
		\langle S_{v}\rangle_{\rho_{ABC}}&\leq& \frac{8(1-q)\lambda_{max}(\overline{\rho_{ABC}(I_{2}\otimes \rho_{ij})})}{q\lambda_{k}(I_{2}\otimes \rho_{ij})}\nonumber \\ 
		&+&\frac{16(1-q)S_{NL}(\rho_{ij})\lambda_{min}(\rho_{ABC})}{q\lambda_{k}(I_{2}\otimes \rho_{ij})} 
		\label{ubs} 
	\end{eqnarray}
	\subsubsection{When $\rho_{ij}$ is not detected by $W_{CHSH}$}
	\noindent When the entangled state $\rho_{ij}$ is not detected by $W_{CHSH}$, the expression of the strength of the non-locality  is given by $S^{New}_{NL}(\rho_{ij})$. Therefore, we can re-write (\ref{snldef}) for the entangled state $\rho_{ij}$ as
	\begin{eqnarray}
		S^{New}_{NL}(\rho_{ij})=r(P^{max}-\frac{3}{4})+(1-r)K,~0\leq r \leq 1
		\label{snlnd}
	\end{eqnarray}
	where $K$ is given by 
	\begin{eqnarray}
		K&=&\frac{Tr[W_{CHSH}\rho_{ij}(\rho_{ij})^{T_{j}}]}{4N(\rho_{ij})}\nonumber 
	\end{eqnarray}
	Here $N(\rho_{ij})$ denote the negativity of the state $\rho_{ij}$ and upper bound of $r$ is given by
	\begin{eqnarray}
		r<\frac{K}{\frac{3}{4}-P^{max}+K}
		\label{qub}
	\end{eqnarray} 
	It can be shown that the quantity $K$ satisfies the inequality
	\begin{eqnarray}
		K\leq \frac{\lambda_{max}(W_{CHSH})\lambda_{max}(\rho_{ij})Tr[W_{CHSH}\rho_{ij}]+Tr[(\rho_{ij}^{T_{j}})^{2}]}{8N(\rho_{ij})} \nonumber\\
		\label{K2}
	\end{eqnarray}
	The proof of the derivation of the inequality (\ref{K2}) is given in the Appendix-II.\\ 
	Using (\ref{K2}) in (\ref{snlnd}), $Tr(W_{CHSH}\rho_{ij})$ may be estimated as
	\begin{eqnarray}
		&&Tr[W_{CHSH}\rho_{ij}]\geq \frac{1}{\lambda_{max}(W_{CHSH})\lambda_{max}(\rho_{ij})}\times\nonumber \\&&\big[\frac{8N(\rho_{ij})(S^{New}_{NL}(\rho_{ij})-r(P^{max}-\frac{3}{4}))}{1-r}-Tr[(\rho_{ij}^{T_{j}})^{2}]\big] \nonumber \\ 
		\label{wrho}
	\end{eqnarray} 
	Using (\ref{wrho}), the inequality (\ref{upperbound}) for the upper bound of $\langle S_{v}\rangle$ reduces to
	\begin{eqnarray}
		\langle S_{v}\rangle_{\rho_{ABC}}&\leq& \frac{2(1-q)}{q\lambda_{k}(I_{2}\otimes \rho_{ij})}
		\biggr[4\lambda_{max}(\overline{\rho_{ABC}(I_{2}\otimes \rho_{ij})})\nonumber \\&-&\frac{\lambda_{min}(\rho_{ABC})\times A_{2}}{\lambda_{max}(W_{CHSH})\lambda_{max}(\rho_{ij})}\biggl]
		\label{ubs1}
	\end{eqnarray}
where $A_{2}= \frac{8N(\rho_{ij})(S^{New}_{NL}(\rho_{ij})-r(P^{max}-\frac{3}{4}))}{1-r}-Tr[(\rho_{ij}^{T_{j}})^{2}]$\\
	The results given by (\ref{ubs}) and (\ref{ubs1}) can be collectively given by the following theorem:\\
	
\noindent	\textbf{Theorem-1b:} The upper bound of the expectation value of the Svetlichny operator $S_{v}$ with respect to any three-qubit state $\rho_{ABC}$ can be expressed in terms of $S_{NL}(\rho_{ij})$ and $S^{New}_{NL}(\rho_{ij})$ as
	\begin{eqnarray}
		(i)~~\langle S_{v}\rangle_{\rho_{ABC}}&\leq& \frac{8(1-q)\lambda_{max}(\overline{\rho_{ABC}(I_{2}\otimes \rho_{ij})})}{q\lambda_{k}(I_{2}\otimes \rho_{ij})}\nonumber \\ 
		&+&\frac{16(1-q)S_{NL}(\rho_{ij})\lambda_{min}(\rho_{ABC})}{q\lambda_{k}(I_{2}\otimes \rho_{ij})}
	\end{eqnarray}
	and
\begin{eqnarray}
(ii)~~\langle S_{v}\rangle_{\rho_{ABC}}&\leq& \frac{2(1-q)}{q\lambda_{k}(I_{2}\otimes \rho_{ij})}
	\biggr[4\lambda_{max}(\overline{\rho_{ABC}(I_{2}\otimes \rho_{ij})})\nonumber \\&-&\frac{\lambda_{min}(\rho_{ABC})\times A_{2}}{\lambda_{max}(W_{CHSH})\lambda_{max}(\rho_{ij})}\biggl]
\end{eqnarray}
where $A_{2}= \frac{8N(\rho_{ij})(S^{New}_{NL}(\rho_{ij})-r(P^{max}-\frac{3}{4}))}{1-r}-Tr[(\rho_{ij}^{T_{j}})^{2}]$
	according as when the entangled state $\rho_{ij}$ does or does not detected by the witness operator $W_{CHSH}$.\\ 
	\section{Detection of genuine three-qubit non-local states}
	\noindent In this section, we will derive conditions to identify whether the given three-qubit state (pure or mixed) is a genuine non-local state. We will use the Svetlichny inequality and the lower and upper bound given in theorem-1a and theorem-1b stated in the previous section, to derive much simpler conditions than the Svetlichny inequality for the detection of genuine non-locality of the three-qubit state. We will show that the genuine non-locality of the  three-qubit state depends on the non-locality of the two-qubit reduced entangled state. The non-locality of two-qubit reduced entangled state $\rho_{ij}$ may be determined by $S_{NL}(\rho_{ij})$ and $S^{New}_{NL}(\rho_{ij})$ accordingly the entangled state $\rho_{ij}$ detected and not detected by the CHSH witness operator $W_{CHSH}$. 
	\subsection{When $\rho_{ij}$ is detected by the witness operator $W_{CHSH}$}
	\noindent In this section, we will derive the condition of non-locality of the three-qubit state described by the density operator $\rho_{ABC}$ when its reduced two-qubit entangled state $\rho_{ij}$ is detected by the witness operator $W_{CHSH}$.\\ 
	
\noindent	\textbf{Theorem-2a:} If any three-qubit state (either pure or mixed) satisfies the Svetlichny inequality and if the reduced two-qubit state of it is detected by the CHSH witness operator  then the operators $A_{l}$ and $B_{l}$ given in (\ref{cons1a}) and (\ref{cons1b}) must be chosen in such a way that the parameter $p$ given by (\ref{cons1a}) satisfies the following inequality\\
	(i) If $\lambda_{min}(\overline{\rho_{ABC}(I_{2}\otimes\rho_{ij})})+ 2S_{NL}(\rho_{ij})\lambda_{max}(\rho_{ABC})> 0$, then 
	\begin{eqnarray}
		0\leq p \leq u_{1}, \text{when}~d_{1}^{(-)}>0  
		\label{thm2a1}
	\end{eqnarray}
 \begin{center}
 	OR
 \end{center}
	\begin{eqnarray}
	l_{1}\leq p \leq 1 ,\text{when}~d_{1}^{(+)}>0 
	\label{thm2aa1}
\end{eqnarray}
	(ii) If $\lambda_{min}(\overline{\rho_{ABC}(I_{2}\otimes\rho_{ij})})+ 2S_{NL}(\rho_{ij})\lambda_{max}(\rho_{ABC}) < 0$, then 
	\begin{eqnarray}
		u_{1}\leq p \leq 1, \text{when}~d_{1}^{(-)}<0 
		\label{thm2a2}
	\end{eqnarray}
 \begin{center}
	OR
\end{center}
\begin{eqnarray}
	0\leq p \leq l_{1}, \text{when}~d_{1}^{(+)}<0  
	\label{thm2aa2}
\end{eqnarray}
	The lower bound $l_{1}$ and upper bound $u_{1}$ are given by
	\begin{eqnarray}
		l_{1}=\frac{2}{d^{(+)}_{1}}&\times&[\lambda_{min}(\overline{\rho_{ABC}(I_{2}\otimes\rho_{ij})})+ 2S_{NL}(\rho_{ij})\times\nonumber\\&&\lambda_{max}(\rho_{ABC})] 
		\label{l1}
	\end{eqnarray}
	\begin{eqnarray}
		u_{1}=\frac{2}{d^{(-)}_{1}}&\times&[\lambda_{min}(\overline{\rho_{ABC}(I_{2}\otimes\rho_{ij})})+ 2S_{NL}(\rho_{ij})\times\nonumber\\&&\lambda_{max}(\rho_{ABC})]
		\label{u1}
	\end{eqnarray}
	where $d^{(+)}_{1}=2[\lambda_{min}(\overline{\rho_{ABC}(I_{2}\otimes\rho_{ij})})+ 2S_{NL}(\rho_{ij})\lambda_{max}(\rho_{ABC})]+\lambda_{max}(I_{2}\otimes\rho_{ij})$ and 
	$d^{(-)}_{1}=2[\lambda_{min}(\overline{\rho_{ABC}(I_{2}\otimes\rho_{ij})})+ 2S_{NL}(\rho_{ij})\lambda_{max}(\rho_{ABC})]-\lambda_{max}(I_{2}\otimes\rho_{ij})$.\\
	\textbf{Proof:} Let us consider a three-qubit state $\rho_{ABC}$ which satisfies the Svetlichny inequality. Therefore, we have 
	\begin{eqnarray}\label{eq2}
		-4 \leq \langle S_{v}\rangle_{\rho_{ABC}} \leq 4
		\label{svin100}
	\end{eqnarray}
	Now, if a three-qubit state $\rho_{ABC}$ satisfies the Svetlichny inequality then our task is to construct the operator $A_{l}$. To accomplish this task, we need to specify the parameter $p$. Thus, recalling the lower bound of the expectation value of the Svetlichny operator $S_{v}$ given in (\ref{lb1}) and using (\ref{svin100}), the restriction on $p$ may be obtained by solving the inequality
	\begin{eqnarray}
		-4 &\leq& \frac{8(1-p)\lambda_{min}(\overline{\rho_{ABC}(I_{2}\otimes \rho_{ij})})}{p\lambda_{max}(I_{2}\otimes \rho_{ij})}\nonumber \\ 
		&+&\frac{16(1-p)\lambda_{max}(\rho_{ABC})S_{NL}(\rho_{ij})}{p\lambda_{max}(I_{2}\otimes \rho_{ij})}\nonumber\\&\leq& 4
		\label{ineq1} 
	\end{eqnarray} 
	Solving the inequality (\ref{ineq1}) for the parameter $p$ while considering all the cases when $\lambda_{min}(\overline{\rho_{ABC}(I_{2}\otimes\rho_{ij})})+ 2S_{NL}(\rho_{ij})\lambda_{max}(\rho_{ABC})>0$, and $\lambda_{min}(\overline{\rho_{ABC}(I_{2}\otimes\rho_{ij})})+ 2S_{NL}(\rho_{ij})\lambda_{max}(\rho_{ABC})<0$, we get the required result. Hence proved.\\
	
	\noindent \textbf{Corollary-1a:} Let us define the quantity $U^{(1)}_{n}=\sqrt{2}[\lambda_{min}(\overline{\rho_{ABC}(I_{2}\otimes\rho_{ij})}) + 2S_{NL}(\rho_{ij})\lambda_{max}(\rho_{ABC})]$, $U^{(1)}_{-}=\sqrt{2}[\lambda_{min}(\overline{\rho_{ABC}(I_{2}\otimes\rho_{ij})})+ 2S_{NL}(\rho_{ij})\lambda_{max}(\rho_{ABC})]\lambda_{max}(I_{2}\otimes\rho_{ij})$, and $U^{(1)}_{+}=\sqrt{2}[\lambda_{min}(\overline{\rho_{ABC}(I_{2}\otimes\rho_{ij})})+ 2S_{NL}(\rho_{ij})\lambda_{max}(\rho_{ABC})]+\lambda_{max}(I_{2}\otimes\rho_{ij})$. If the parameter $p$ violate (\ref{thm2aa1}) and (\ref{thm2a2}) for some three-qubit (pure or mixed) state $\rho_{ABC}$ i.e. if it satisfies the inequality 
	\begin{eqnarray}
	\frac{U^{(1)}_{n}}{U^{(1)}_{+}} < p < l_{1}
		\label{cor21}
	\end{eqnarray}
when  $\lambda_{min}(\overline{\rho_{ABC}(I_{2}\otimes\rho_{ij})})+ 2S_{NL}(\rho_{ij})\lambda_{max}(\rho_{ABC})> 0$
\begin{center}

	OR
\end{center}
	\begin{eqnarray}
\frac{U^{(1)}_{n}}{U^{(1)}_{-}} < p < u_{1}
	\label{cor21a}
\end{eqnarray}
when $\lambda_{min}(\overline{\rho_{ABC}(I_{2}\otimes\rho_{ij})})+ 2S_{NL}(\rho_{ij})\lambda_{max}(\rho_{ABC})< 0$
	then the state $\rho_{ABC}$ violates the Svetlichny inequality and thus exhibits the genuine non-locality.\\
\textbf{Note-1:} We should note here that the expression of $\frac{U^{(1)}_{n}}{U^{(1)}_{+}}$ and $\frac{U^{(1)}_{n}}{U^{(1)}_{-}}$ has been obtained by using the upper limit of $\langle S_{v} \rangle_{\rho_{ABC}}$ i.e. $\langle S_{v} \rangle_{\rho_{ABC}} \leq 4\sqrt{2}$.\\
	
\noindent	\textbf{Theorem-2b:} If any three-qubit state (either pure or mixed) satisfies the Svetlichny inequality and if the reduced two-qubit state of it is detected by the CHSH witness operator then the  operators $A_{u}$ and $B_{u}$ given in (\ref{cons2a}) and (\ref{cons2b}) must be chosen in such a way that the parameter $q$ given by (\ref{cons2a}) satisfies the inequality
	\begin{eqnarray}
		l_{2}\leq q \leq 1 
		\label{thm2b}
	\end{eqnarray}
	The lower bound $l_{2}$ is given by
	\begin{eqnarray}
		l_{2}=\frac{2}{d^{(+)}_{2}}&\times& [\lambda_{max}(\overline{\rho_{ABC}(I_{2}\otimes\rho_{ij})}) + 2S_{NL}(\rho_{ij})\nonumber\\&&\lambda_{min}(\rho_{ABC})] 
		\label{l2}
	\end{eqnarray}
	where $d^{(+)}_{2}=2[\lambda_{max}(\overline{\rho_{ABC}(I_{2}\otimes\rho_{ij})})+ 2S_{NL}(\rho_{ij})\lambda_{min}(\rho_{ABC})]+\lambda_{k}(I_{2}\otimes\rho_{ij})$.\\
	Proof of $\textbf{theorem-2b}$ is given in $Appendix-III$.\\
	
	\noindent \textbf{Corollary-1b:} Let us define the quantity $U^{(2)}_{n}=\sqrt{2}[\lambda_{max}(\overline{\rho_{ABC}(I_{2}\otimes\rho_{ij})}) + 2S_{NL}(\rho_{ij})\lambda_{min}(\rho_{ABC})]$ and $U^{(2)}_{d}=\sqrt{2}[\lambda_{max}(\overline{\rho_{ABC}(I_{2}\otimes\rho_{ij})})+ 2S_{NL}(\rho_{ij})\lambda_{min}(\rho_{ABC})]+\lambda_{k}(I_{2}\otimes\rho_{ij})$. If the parameter $q$ violates the inequality given in (\ref{thm2b}) for some three-qubit (pure or mixed) state $\rho_{ABC}$ i.e. if it satisfies the inequality 
	\begin{eqnarray}
		U^{(2)}\equiv\frac{U^{(2)}_{n}}{U^{(2)}_{d}} < q < l_{2} 
		\label{cor2b}
	\end{eqnarray}
	then the state $\rho_{ABC}$ violates the Svetlichny inequality and thus exhibits the genuine non-locality.\\
	 
\noindent	\textbf{Result-2:} If any three-qubit state (either pure or mixed) satisfies the Svetlichny inequality then the Svetlichny operator also satisfies the inequality
	\begin{eqnarray}
		S_{v}^{(1)} \leq \langle S_{v} \rangle_{\rho_{ABC}} \leq S_{v}^{(2)} 
		\label{thm3}
	\end{eqnarray}
	where $S_{v}^{(1)}$ and $S_{v}^{(2)}$ are given by
	\begin{eqnarray}
		S_{v}^{(1)}&=&\frac{8(1-p)}{p\lambda_{max}(I_{2}\otimes \rho_{ij})}[\lambda_{min}(\overline{\rho_{ABC}(I_{2}\otimes \rho_{ij})})\nonumber\\&+&2\lambda_{max}(\rho_{ABC})S_{NL}(\rho_{ij})]
		\label{slb1}
	\end{eqnarray}
	\begin{eqnarray}
		S_{v}^{(2)}&=&\frac{8(1-q)}{q\lambda_{k}(I_{2}\otimes \rho_{ij})}[\lambda_{max}(\overline{\rho_{ABC}(I_{2}\otimes \rho_{ij})})\nonumber\\&+&2\lambda_{min}(\rho_{ABC})S_{NL}(\rho_{ij})]
		\label{sub1}
	\end{eqnarray}
	The two parameters $p$ and $q$ satisfies the inequality (\ref{thm2a1}),  (\ref{thm2aa1}), (\ref{thm2a2}), (\ref{thm2aa2}) and (\ref{thm2b}).\\
	 
	\noindent \textbf{Corollary-1c:} If any three-qubit state (either pure or mixed) violate the inequality (\ref{thm3}) and if $p$ and $q$ satisfies the inequality (\ref{cor21}), (\ref{cor21a}) and (\ref{cor2b}) then the given three-qubit state exhibit genuine non-locality. In other words, for any three-qubit state (either pure or mixed) described by the density operator $\rho_{ABC}$ if
	\begin{eqnarray}
		\langle S_{v} \rangle_{\rho_{ABC}} < S_{v}^{(1)},~~\langle S_{v} \rangle_{\rho_{ABC}}> S_{v}^{(2)}
		\label{violcond1}
	\end{eqnarray}
	then $\rho_{ABC}$ exhibit genuine non-locality.
	\subsection{When $\rho_{ij}$ is not detected by the witness operator $W_{CHSH}$}
	\noindent In this section, we will derive the condition of the non-locality of three-qubit state described by the density operator $\rho_{ABC}$ when its reduced two-qubit entangled state $\rho_{ij}$ is not detected by the Witness operator $W_{CHSH}$.\\ 
	
	\noindent \textbf{Theorem-3a:} If any three-qubit state (either pure or mixed) satisfies the Svetlichny inequality and if the reduced two-qubit state of it is not detected by the CHSH witness operator then the operators $A_{l}$ and $B_{l}$ given in (\ref{cons1a}) and (\ref{cons1b}) must be chosen in such a way that the parameter $p$ given by (\ref{cons1a}) satisfies the following inequality\\
	\begin{eqnarray}
		l_{3}\leq p \leq 1 
		\label{thm3a}
	\end{eqnarray}
	The bound $l_{3}$ is given by
	\begin{eqnarray}
		l_{3}= \frac{2H}{2H-\lambda_{max}(I_{2}\otimes \rho_{ij})}
		\label{l3}
	\end{eqnarray}
	where $H= \lambda_{min}(\overline{\rho_{ABC}(I_{2}\otimes \rho_{ij})})-(\frac{S^{New}_{NL}(\rho_{ij})-r(P^{max}-\frac{3}{4})}{(1-r)\lambda_{min}[(\rho_{ij}^{T_{j}})^{2}]})\times (N(\rho_{ij})\lambda_{max}(\rho_{ij}^{T_{j}})\lambda_{max}(\rho_{ABC}))$.\\
	Proof of $\textbf{theorem-3a}$ is given in $Appendix-IV$.\\
	
	\noindent \textbf{Corollary-2a:} If the parameter $p$ violate the inequality given in (\ref{thm3a}) for some three-qubit (pure or mixed) state $\rho_{ABC}$ i.e. if $p$ satisfies the inequality 
	\begin{eqnarray}
		U^{(3)}\equiv\frac{\sqrt{2}H}{\sqrt{2}H-\lambda_{max}(I_{2}\otimes \rho_{ij})} < p < l_{3} 
		\label{cor}
	\end{eqnarray}
	then the three-qubit state $\rho_{ABC}$ violates the Svetlichny inequality and thus exhibits the genuine non-locality.\\
	
	\noindent \textbf{Theorem-3b:} If any three-qubit state (either pure or mixed) satisfies the Svetlichny inequality and if the reduced two-qubit state of it is not detected by the CHSH witness operator then the operators $A_{u}$ and $B_{u}$ given in (\ref{cons2a}) and (\ref{cons2b}) must be chosen in such a way that the parameter $q$ given by (\ref{cons2a}) satisfies the inequality
	\begin{eqnarray}
		l_{4}\leq q \leq 1 
		\label{thm4b}
	\end{eqnarray}
	The bounds $l_{4}$ is given by
	\begin{eqnarray}
		l_{4}=\frac{F}{F+ 2\lambda_{k}(I_{2}\otimes \rho_{ij})}
		\label{l4}
	\end{eqnarray}
	where $F=(4\lambda_{max}(\overline{\rho_{ABC}(I_{2}\otimes \rho_{ij})})-\frac{\lambda_{min}(\rho_{ABC})}{\lambda_{max}(W_{CHSH})\lambda_{max}(\rho_{ij})}(\frac{8N(\rho_{ij})(S^{New}_{NL}(\rho_{ij})-r(P^{max}-\frac{3}{4}))}{1-r}-Tr[(\rho_{ij}^{T_{j}})^{2}]))$.\\
	Proof of $\textbf{theorem-3b}$ is given in $Appendix-V$.\\
	
	\noindent \textbf{Corollary-2b:} If the parameter $q$ violates the inequality given in (\ref{thm4b}) for some three-qubit (pure or mixed) state $\rho_{ABC}$ i.e. if it satisfies the inequality 
	\begin{eqnarray}
		U^{(4)}\equiv\frac{F}{F+ 2\sqrt{2}\lambda_{k}(I_{2}\otimes \rho_{ij})}< q < l_{4} 
		\label{cor200}
	\end{eqnarray}
	then the three-qubit state violates the Svetlichny inequality and thus exhibits the genuine non-locality.\\
	
	\noindent \textbf{Result-3:} If any three-qubit state (either pure or mixed) satisfies the Svetlichny inequality and if $p$ and $q$ are given by (\ref{thm3a}) and (\ref{thm4b}) then the Svetlichny operator also satisfies the inequality
	\begin{eqnarray}
		S_{v}^{(3)} \leq \langle S_{v} \rangle_{\rho_{ABC}} \leq S_{v}^{(4)} 
		\label{thm5}
	\end{eqnarray}
	where $S_{v}^{(3)}$ and $S_{v}^{(4)}$ are given by
	\begin{eqnarray}
		S_{v}^{(3)}&=& \frac{8(1-p)}{p\lambda_{max}(I_{2}\otimes \rho_{ij})}\biggl[\lambda_{min}(\overline{\rho_{ABC}(I_{2}\otimes \rho_{ij})})-\nonumber\\&&\frac{\big(S^{New}_{NL}(\rho_{ij})-r(P^{max}-\frac{3}{4})\times A_{1}\big)}{(1-r)\lambda_{min}[(\rho_{ij}^{T_{j}})^{2}]}\biggr]
		\label{slb3}
	\end{eqnarray}
	\begin{eqnarray}
		S_{v}^{(4)}&=& \frac{2(1-q)}{q\lambda_{k}(I_{2}\otimes \rho_{ij})}
		\biggl[4\lambda_{max}(\overline{\rho_{ABC}(I_{2}\otimes \rho_{ij})})\nonumber \\&-&\frac{\lambda_{min}(\rho_{ABC})\times A_{2}}{\lambda_{max}(W_{CHSH})\lambda_{max}(\rho_{ij})}\biggr]
		\label{sub4}
	\end{eqnarray}
where $A_{1}=(N(\rho_{ij})\lambda_{max}(\rho_{ij}^{T_{j}})\lambda_{max}(\rho_{ABC}))$ and $A_{2}= \frac{8N(\rho_{ij})(S^{New}_{NL}(\rho_{ij})-r(P^{max}-\frac{3}{4}))}{1-r}-Tr[(\rho_{ij}^{T_{j}})^{2}]$.\\

\noindent	\textbf{Corollary-2c:} If any three-qubit state (either pure or mixed) violates the inequality (\ref{thm5}) and if $p$ and $q$ satisfy the inequality given by (\ref{cor}) and (\ref{cor200}) then the given three-qubit state exhibit genuine non-locality. In other words, for any three-qubit state (either pure or mixed) described by the density operator $\rho_{ABC}$ if
	\begin{eqnarray}
		\langle S_{v} \rangle_{\rho_{ABC}} < S_{v}^{(3)},~~\langle S_{v} \rangle_{\rho_{ABC}}> S_{v}^{(4)}
		\label{violcond}
	\end{eqnarray}
	then $\rho_{ABC}$ exhibit genuine non-locality.
	
	\section{Illustrations}
	\noindent We are now in a position to illustrate our scheme of finding the genuine non-locality of a given three-qubit state (pure or mixed) with a few examples.
	\subsection{When the reduced two-qubit state $\rho_{ij}$ is detected by the CHSH witness operator}
	\noindent In this section, we will illustrate our results given in (\ref{violcond1}) with the help of the following two examples of three-qubit states for which the reduced two-qubit state is detected by the CHSH witness operator: (i) A pure three-qubit state belong to W class and (ii) A mixed three-qubit state which may be taken as a convex combination of GHZ state and two other states belong to W class. 
	\subsubsection{A pure three-qubit W class of state}
\noindent	Let us consider a pure three-qubit state of the form 
	\begin{eqnarray}
		|\psi^{(1)}\rangle_{ABC}&=& \lambda_{0}|000\rangle +0.3|101\rangle \nonumber \\&+& \sqrt{0.91-\lambda_{0}^{2}} |110\rangle 
		\label{pstate1}
	\end{eqnarray}
	where the state parameter $\lambda_{0}\in [0,0.953939]$.\\
	The pure state described by the density operator $\rho_{ABC}^{(1)}= |\psi^{1}\rangle_{ABC}\langle\psi^{1}|$ is an entangled state and also we have
	\begin{eqnarray}
		\lambda_{max}(\rho_{ABC}^{(1)})=1,~~\lambda_{min}(\rho_{ABC}^{(1)})=0
		\label{eigval1} 
	\end{eqnarray}
	Tracing out system B from the three-qubit state $\rho_{ABC}^{(1)}$, the reduced state $\rho^{(1)}_{AC}$ is given by
	\begin{eqnarray}
		\rho^{(1)}_{AC}=\begin{pmatrix} 
			\lambda_{0}^{2} & 0& 0&0.3\lambda_{0} \\
			0 & 0 & 0 & 0\\
			0 & 0 & 0.91-\lambda_{0}^{2} & 0\\
			0.3\lambda_{0} & 0& 0& 0.09\\
		\end{pmatrix}
	\end{eqnarray}
	The state $\rho^{(1)}_{AC}$ is an entangled state as there exist a witness operator $W_{CHSH}^{(xz)}(=2I-B_{CHSH}^{(xz)})$ that detect it. The CHSH witness operator $B_{CHSH}^{(xz)}$ is given by (\ref{bchsh}). This is clear from the following fact 
	\begin{eqnarray}
		Tr[W_{CHSH}^{(xz)}\rho^{(1)}_{AC}]&=&3.15966-0.848528\lambda_{0}-2.82843\lambda_{0}^{2}\nonumber\\&<& 0,~~\text{for}~~\lambda_{0}\in [0.91753,0.953939]
	\end{eqnarray}
	Since the two-qubit state $\rho^{(1)}_{AC}$ is an entangled state and it is detected by $W_{CHSH}^{(xz)}$ so the strength of its non-locality may be measured by $S_{NL}(\rho^{(1)}_{AC})$. It is then given by
	\begin{eqnarray}
		S_{NL}(\rho^{(1)}_{AC}) &=&\frac{-Tr[W_{CHSH}^{(xz)}\rho^{(1)}_{AC}]}{8}\nonumber\\
		&\in&~~[0,0.030],~\text{for}~\lambda_{0}\in [0.917,0.953]
		\label{snlinfo1}
	\end{eqnarray}
	Further, we can calculate the following using the three-qubit state $\rho_{ABC}^{(1)}$ and the reduced two-qubit state $\rho^{(1)}_{AC}$
	\begin{eqnarray}
		&&\lambda_{max}(I_{2}\otimes \rho_{AC}^{(1)})=0.09+\lambda_{0}^{2}\nonumber\\&&
		\lambda_{min}(\overline{\rho_{ABC}^{(1)}(I_{2}\otimes \rho_{AC}^{(1)})})= \frac{\lambda_{0}^{4}-\lambda_{0}^{3}}{2}+\frac{9(\lambda_{0}^{2}-\lambda_{0})}{200} \nonumber\\&&
		\lambda_{max}(\overline{\rho_{ABC}^{(1)}(I_{2}\otimes \rho_{AC}^{(1)})})=\frac{\lambda_{0}^{4}+\lambda_{0}^{3}}{2}+\frac{9(\lambda_{0}^{2}+\lambda_{0})}{200}\nonumber\\&&
		\lambda_{k}(I_{2}\otimes \rho_{AC}^{(1)})=0.91-\lambda_{0}^{2}
		\label{info1}
	\end{eqnarray}
	Also, the range of $p$ and $q$ 
	are given by
	\begin{eqnarray}
	0 < p <  0.07
		\label{ranp1}
	\end{eqnarray}
	\begin{eqnarray}
		0.93 <  q <  1
		\label{ranq1}
	\end{eqnarray}
	Using the information given in (\ref{eigval1}), (\ref{snlinfo1}), (\ref{info1}), (\ref{ranp1}), and (\ref{ranq1}), the value of the expression of $S_{v}^{(1)}$ and $S_{v}^{(2)}$ can be calculated for the three-qubit state $\rho_{ABC}^{(1)}$ and they are tabulated in the Table-\ref{t1}.

	\subsubsection{A mixed three-qubit state: Combination of GHZ state and two W class of states}
\noindent	Let us consider a mixed three-qubit state of the form\cite{jung} 
	\begin{eqnarray}
		\rho^{(2)}_{ABC}&=& 0.2 |GHZ\rangle\langle GHZ|+t|W_{1}\rangle\langle W_{1}|\nonumber\\&+&(0.8-t)|W_{2}\rangle\langle W_{2}|, ~~ t\in [0,0.8] 
		\label{mstate2}
	\end{eqnarray}
	where $|GHZ\rangle=\frac{1}{\sqrt{2}}(|000\rangle+|111\rangle), |W_{1}\rangle=\frac{1}{\sqrt{3}}(|001\rangle+|010\rangle+|100\rangle), |W_{2}\rangle=\frac{1}{\sqrt{3}}(|110\rangle+|101\rangle+|011\rangle)$.\\
	The mixed three-qubit state described by the density operator $\rho_{ABC}^{(2)}$ is an entangled state when $t\in [0,0.8]$ and also we have
	\begin{eqnarray}
		\lambda_{max}(\rho_{ABC}^{(2)})=t,~~\lambda_{min}(\rho_{ABC}^{(2)})=0
		\label{eigval2} 
	\end{eqnarray}
	Tracing out system A from the three-qubit state $\rho_{ABC}^{(2)}$, the reduced state $\rho^{(2)}_{BC}$ is given by
	\begin{eqnarray}
		\rho^{(2)}_{BC}=\begin{pmatrix} 
			\frac{0.6+2t}{6} & 0& 0 & 0 \\
			0 & \frac{0.8}{3} & \frac{0.8}{3} & 0\\
			0 & \frac{0.8}{3} & \frac{0.8}{3} & 0\\
			0 & 0 & 0 & \frac{2.2-2t}{6}\\
		\end{pmatrix}
	\end{eqnarray}
	The state $\rho^{(2)}_{BC}$ is an entangled state for $t\in [0.5,0.8]$.\\
	Let us now consider the witness operator $W_{CHSH}$, which is given by  
	\begin{eqnarray}
		W_{CHSH}&=&2I\otimes I-A_{0}\otimes B_{0}+A_{0}\otimes B_{1}\nonumber\\&&-A_{1}\otimes B_{0}-A_{1}\otimes B_{1} 
		\label{witchsh2}
	\end{eqnarray}
	where the Hermitian operators $A_{0}$, $A_{1}$, $B_{0}$, $B_{1}$ are given by
	\begin{eqnarray}
		A_{0}&=&\sigma_{x}\nonumber \\
		A_{1}&=&\sigma_{y}\nonumber \\ 
		B_{0}&=&0.95\sigma_{x}+0.95\sigma_{y}+0.447\sigma_{z}\nonumber \\
		B_{1}&=&-0.95\sigma_{x}+0.95\sigma_{y}+0.447\sigma_{z}
	\end{eqnarray}
	The expectation value of $W_{CHSH}$ with respect to the two-qubit state $\rho_{BC}^{(2)}$ can be calculated as
	\begin{eqnarray}
		Tr[W_{CHSH}\rho_{BC}^{(2)}]= -0.0266667 < 0  
	\end{eqnarray}
	Therefore, the two-qubit state $\rho_{BC}^{(2)}$ is detected by witness operator $W_{CHSH}$. Thus, the strength of its non-locality may be measured by $S_{NL}(\rho^{(2)}_{BC})$, which is given by 
	\begin{eqnarray}
		S_{NL}(\rho^{(2)}_{BC}) &=&\frac{-Tr[W_{CHSH}\rho^{(2)}_{BC}]}{8}\nonumber\\
		&=& 0.00333,~\text{for}~ t \in [0.5,0.8]
		\label{snlinfo2}
	\end{eqnarray}
	Further, we are now in a position to calculate the value of the following expressions involving the three-qubit state $\rho_{ABC}^{(2)}$ and the reduced two-qubit state $\rho^{(2)}_{BC}$
	\begin{eqnarray}
		&&\lambda_{max}(I_{2}\otimes \rho_{BC}^{(2)})=0.5333\nonumber\\&&
		\lambda_{k}(I_{2}\otimes \rho_{BC}^{(2)})=0.333(1.1-t)
		\label{info2}
	\end{eqnarray}
	Also, the range of $p$ and $q$ 
	are given by
	\begin{eqnarray}
		0 < p <  0.05
		\label{ranp2}
	\end{eqnarray}
	\begin{eqnarray}
	0.34 <  q < 0.37
		\label{ranq2}
	\end{eqnarray}
	Using the information given in (\ref{eigval2}), (\ref{snlinfo2}), (\ref{info2}), (\ref{ranp2}), and (\ref{ranq2}), the value of the expression of $S_{v}^{(1)}$ and $S_{v}^{(2)}$ can be tabulated for the three-qubit state $\rho_{ABC}^{(2)}$ in Table-\ref{t2}.\\

	\subsection{When the reduced two-qubit state $\rho_{ij}$ is not detected by $W_{CHSH}$}
	\noindent In this section, we have considered three examples of three-qubit states in which the reduced two-qubit states are not detected by CHSH witness operator $W_{CHSH}$. The three examples are given in the following form: (i) A pure three-qubit state which belong to GHZ class (ii) A mixed state which may be taken as a convex combination of three-qubit GHZ and W state and (iii) A mixed state which may be taken as a convex combination of three-qubit maximally mixed state and W state.
	\subsubsection{A pure three-qubit GHZ class of state: Maximal Slice State}
	\noindent Let us consider a pure three-qubit GHZ class of state, which can be taken in the form\cite{carteret} 
	\begin{eqnarray}
		|\psi^{(3)}\rangle_{ABC}&=&\frac{1}{\sqrt{2}} (|000\rangle_{ABC} +Cos\theta|110\rangle_{ABC}\nonumber\\&+&Sin\theta |111\rangle_{ABC}),~\theta \in [0,\frac{\pi}{2}]
	\end{eqnarray}
	The pure state described by the density operator $\rho_{ABC}^{(3)}= |\psi^{(3)}\rangle_{ABC}\langle\psi^{(3)}|$ is an entangled state for $\theta \in (0,\frac{\pi}{2})$.\\
	Also, for the state $\rho^{(3)}_{ABC}$, we have
	\begin{eqnarray}
		\lambda_{max}(\rho_{ABC}^{(3)})=1,~~\lambda_{min}(\rho_{ABC}^{(3)})=0
		\label{eigval3} 
	\end{eqnarray}
	Tracing out system A from the three-qubit state $\rho_{ABC}^{(3)}$, the reduced two-qubit state $\rho^{(3)}_{BC}$ is given by
	\begin{eqnarray}
		\rho^{(3)}_{BC}=\begin{pmatrix} 
			\frac{1}{2} & 0& 0&Cos\theta\\
			0 & 0 & 0 & 0\\
			0 & 0 & 0 & 0\\
			Cos\theta & 0& 0&\frac{1}{2} \\
		\end{pmatrix}
	\end{eqnarray}
	It can be easily verified that $\rho^{(3)}_{BC}$ is an entangled state for the state parameter $\theta\in [1.05,\frac{\Pi}{2}]$. Thus there must exist a witness operator that may detect $\rho^{(3)}_{BC}$ as an entangled state. But, in this example, our task is to show that even if some witness operator does not detect the reduced two-qubit entangled state then also we are able to detect the non-locality of the three-qubit state described by the density operator $\rho^{(3)}_{BC}$.\\
	To serve our purpose, we find here a witness operator $W_{CHSH}^{(xy)}=2I-B^{(xy)}_{CHSH}$, whose expectation value with respect to the state $\rho^{(3)}_{BC}$ is given by
	$Tr[W_{CHSH}^{(xy)}\rho^{(3)}_{BC}]=2>0$. Thus, the CHSH witness operator $W_{CHSH}^{(xy)}$ does not detect $\rho^{(3)}_{BC}$ as an entangled state. Since the two-qubit state $\rho^{(3)}_{BC}$ is an entangled state and it is not detected by $W_{CHSH}^{(xy)}$ so the strength of its non-locality may be measured by $S_{NL}^{New}(\rho^{(3)}_{BC})$. Using (\ref{snldef}) and (\ref{qub}), we can calculate the range of $S_{NL}^{New}(\rho^{(3)}_{BC})$ and $r$. Therefore, we have 
	\begin{eqnarray}
		S_{NL}^{New}(\rho^{(3)}_{BC})~~\in~~[0.05,1.5],~\theta~~\in~~[\frac{147\pi}{440},\frac{\pi}{2}]
		\label{ex3snl}
	\end{eqnarray}
	and
	\begin{eqnarray}
		r<[0.5,1],~\theta~~\in~~[\frac{147\pi}{440},\frac{\pi}{2}]
		\label{ex3r}
	\end{eqnarray}
	Further, we can calculate the following using the three-qubit state $\rho_{ABC}^{(3)}$ and the reduced two-qubit state $\rho^{(3)}_{BC}$
	\begin{eqnarray}
		&&\lambda_{max}(I_{2}\otimes \rho_{BC}^{(3)})=\frac{1+2Cos\theta}{2}\nonumber\\&&
		\lambda_{min}[\overline{\rho_{ABC}^{(3)}(I_{2}\otimes \rho_{BC}^{(3)})}]=\nonumber\\&&\frac{3-Cos2\theta-2\sqrt{8+3Cos2\theta-Cos4\theta}}{16}\nonumber\\&&
		\lambda_{max}[\overline{\rho_{ABC}^{(3)}(I_{2}\otimes \rho_{BC}^{(3)})}]=\nonumber\\&&\frac{3-Cos2\theta+2\sqrt{8+3Cos2\theta-Cos4\theta}}{16}\nonumber\\&&
		\lambda_{k}(I_{2}\otimes \rho_{BC}^{(3)})=\frac{1-2Cos\theta}{2}\nonumber\\&&
		Tr[W^{(xy)}_{CHSH}\rho_{BC}^{(3)}(\rho_{BC}^{(3)})^{T_{C}}]=1\nonumber\\&&
		\lambda_{min}[((\rho_{BC}^{(3)})^{T_{C}})^{2}]= Cos^{2}\theta \nonumber\\&& \lambda_{max}((\rho_{BC}^{(3)})^{T_{C}})= 0.5 
		\label{info3}
	\end{eqnarray}
	Moreover, the range of $p$ and $q$ 
	are given by
	\begin{eqnarray}
	0.75 < p <  1
		\label{ranp3}
	\end{eqnarray}
	\begin{eqnarray}
	0.59 <  q < 1
		\label{ranq3}
	\end{eqnarray}
	Using the information given in (\ref{eigval3}), (\ref{info3}), (\ref{ranp3}), and (\ref{ranq3}), the value of the expression of $S_{v}^{(3)}$ and $S_{v}^{(4)}$ can be calculated for the three-qubit state $\rho_{ABC}^{(3)}$ and they are tabulated in the Table-\ref{t3}.

	\subsubsection{A three-qubit mixed state: A convex combination of three-qubit W state and a state belong to GHZ class}
	\noindent Let us take a mixed three-qubit state of the form 
	\begin{eqnarray}
		\rho^{(4)}_{ABC}&=& p_{s}|GHZ\rangle\langle GHZ|+(1-p_{s})|W\rangle\langle W|,\\ &&
		p_{s} \in [0,1]\nonumber
		\label{mstate4}
	\end{eqnarray}
	where $|GHZ\rangle=\frac{1}{\sqrt{2}}(|010\rangle+|101\rangle), |W\rangle=\frac{1}{\sqrt{3}}(|001\rangle+|010\rangle+|100\rangle)$.\\
	The mixed three-qubit state described by the density operator $\rho_{ABC}^{(4)}$ is an entangled state when $p_{s}\in [0.4,0.9]$ and also we have
	\begin{eqnarray}
		&&\lambda_{max}(\rho_{ABC}^{(4)})=\frac{3+\sqrt{3}\sqrt{3-10p_{s}+10p_{s}^{2}}}{6},\nonumber\\&&\lambda_{min}(\rho_{ABC}^{(4)})=0
		\label{eigval4} 
	\end{eqnarray}
	Tracing out system A from the three-qubit state $\rho_{ABC}^{(4)}$, the reduced state $\rho^{(4)}_{BC}$ is given by 
	\begin{eqnarray}
		\rho^{(4)}_{BC}=\begin{pmatrix} 
			\frac{1-p_{s}}{3} & 0& 0&0 \\
			0 & \frac{p_{s}}{2}+\frac{1-p_{s}}{3} & \frac{1-p_{s}}{3} & 0\\
			0 & \frac{1-p_{s}}{3} & \frac{p_{s}}{2}+\frac{1-p_{s}}{3} & 0\\
			0 & 0& 0& 0\\
		\end{pmatrix}
	\end{eqnarray}
	$\rho_{BC}^{(4)}$ is an entangled state for  $p_{s}\in [0.4,0.9]$. Also we have
	\begin{eqnarray}
		Tr[W_{CHSH}^{(xy)}\rho^{(4)}_{BC}]&=&\frac{2(3-2\sqrt{2}+2\sqrt{2}p_{s})}{3}>0,\\ &&
		0.4\leq p_{s} \leq 0.9 \nonumber
	\end{eqnarray}
	In this example also, we find that the same CHSH witness operator $W_{CHSH}^{(xy)}$ given in the previous example, is not able to detect the entangled state $\rho^{(4)}_{BC}$.  The strength of the non-locality of $\rho^{(4)}_{BC}$ may be measured by $S_{NL}^{New}(\rho^{(4)}_{BC})$ using (\ref{snldef}). Therefore, $S_{NL}^{New}(\rho^{(4)}_{BC})$ may be calculated as
	\begin{eqnarray}
		S_{NL}^{New}(\rho^{(4)}_{BC})~~\in~~[0.04,1.91628], ~p_{s}~\in~[0.4,0.9]
	\end{eqnarray}
	and the parameter $r$ is given by
	\begin{eqnarray}
		r<[0.59,1], ~p_{s}~\in~[0.4,0.9]
	\end{eqnarray}
	Further, we can calculate the following using the three-qubit state $\rho_{ABC}^{(4)}$ and the reduced two-qubit state $\rho^{(4)}_{BC}$
	\begin{eqnarray}
		&&\lambda_{max}(I_{2}\otimes \rho_{BC}^{(4)})=\frac{4-p_{s}}{6}\nonumber\\&&
		\lambda_{k}(I_{2}\otimes \rho_{BC}^{(4)})=\frac{1-p_{s}}{3}\nonumber\\&&
		Tr[W^{(xy)}_{CHSH}\rho_{BC}^{(4)}(\rho_{BC}^{(4)})^{T_{C}}]\nonumber\\&=&\frac{6-4\sqrt{2}+2\sqrt{2}p_{s}+(3+2\sqrt{2})p_{s}^{2}}{9}\nonumber\\&&
		\lambda_{min}[((\rho_{BC}^{(4)})^{T_{C}})^{2}]=  \nonumber\\&& \frac{3-6p_{s}+3p_{s}^{2}-\sqrt{5}\sqrt{1-4p_{s}+6p_{s}^{2}-4p_{s}^{3}+p_{s}^{(4)}}}{18} \nonumber\\&& \lambda_{max}((\rho_{BC}^{(4)})^{T_{C}})= \frac{2+p_{s}}{6}
		\label{info4}
	\end{eqnarray}
	Also, the range of $p$ in terms of state parameter $p_{s}$ is given by
	\begin{eqnarray}
	\frac{\sqrt{2}H}{\sqrt{2}H-\frac{4-p_{s}}{6}} < p <  \frac{2H}{2H-\frac{4-p_{s}}{6}}
		\label{ranp4}
	\end{eqnarray}
	The range of $q$ in terms of state parameter $p_{s}$ is given by
\begin{eqnarray}
	\frac{F}{F+2\sqrt{2}\frac{1-p_{s}}{3}} < q <  \frac{F}{F-2\sqrt{2}\frac{1-p_{s}}{3}}
		\label{ranq4}
	\end{eqnarray}
	where $F$ and $H$ given in the previous section can be calculated using the information given in (\ref{info4}).\\
	Using the information given in (\ref{eigval4}), (\ref{info4}), (\ref{ranp4}), and (\ref{ranq4}), the value of the expression of $S_{v}^{(3)}$ and $S_{v}^{(4)}$ can be calculated for the three-qubit state $\rho_{ABC}^{(4)}$ and they are tabulated in the Table-\ref{t4}.

	\subsubsection{A three-qubit mixed State: A convex combination of maximally mixed state and W state}
	Let us consider a mixed three-qubit state of the form \cite{chen} 
	\begin{eqnarray}
		\rho^{(5)}_{ABC}=  \frac{1-p_{s}}{8}I_{8}+p_{s}|W\rangle_{ABC}\langle W|, p_{s}\in(0.816,1]
	\end{eqnarray}
	where $I_{8}$ denote the maximally mixed state represented by the Identity matrix and $|W\rangle=\frac{1}{\sqrt{3}}(|001\rangle+|010\rangle+|100\rangle)$.\\
	The mixed three-qubit state described by the density operator $\rho_{ABC}^{(5)}$ is an entangled state when $p_{s}\in (0.816,1]$ and also we have
	\begin{eqnarray}
		\lambda_{max}(\rho_{ABC}^{(5)})=\frac{1+7p_{s}}{8},~\lambda_{min}(\rho_{ABC}^{(5)})=\frac{1-p_{s}}{8}
		\label{eigval5} 
	\end{eqnarray}
	Taking partial trace over the system A, the three-qubit state $\rho_{ABC}^{(5)}$ reduces to the two-qubit state described by the density operator $\rho^{(5)}_{BC}$, which is given by 
	\begin{eqnarray}
		\rho^{(5)}_{BC}=\begin{pmatrix} 
			\frac{p_{s}}{3}+\frac{1-p_{s}}{4} & 0& 0&0 \\
			0 & \frac{p_{s}}{3}+\frac{1-p_{s}}{4} & \frac{p_{s}}{3} & 0\\
			0 & \frac{p_{s}}{3} & \frac{p_{s}}{3}+\frac{1-p_{s}}{4} & 0\\
			0 & 0& 0& \frac{1-p_{s}}{4}
		\end{pmatrix}
	\end{eqnarray}
	where $0.816 < p_{s} \leq 1$.\\
	$\rho_{BC}^{(5)}$ is an entangled state for  $p_{s}\in (0.816,1]$ but we find that
	\begin{eqnarray}
		Tr[W_{CHSH}^{(xy)}\rho^{(5)}_{BC}]&=&2-\frac{4\sqrt{2}p_{s}}{3}>0
		\label{ex5wit}
	\end{eqnarray}
	where $0.816 < p_{s} \leq 1$.\\
	(\ref{ex5wit}) implies that the CHSH witness operator does not detect the entangled state $\rho^{(5)}_{BC}$. 
	The strength of the non-locality of the two-qubit reduced state may be measured by $S^{New}_{NL}(\rho^{(5)}_{BC})$. The strength $S^{New}_{NL}(\rho^{(5)}_{BC})$ and the parameter $r$ is given by
	\begin{eqnarray}
		&&S^{New}_{NL}(\rho^{(5)}_{BC}) \in [0.54124, 0.5484], ~0.816 < p_{s} \leq 1\nonumber\\&&
		r \in [0.61,0.69], ~0.816 < p_{s} \leq 1
	\end{eqnarray}

	\noindent Further, we can now calculate the following values of the expressions using the three-qubit state $\rho_{ABC}^{(5)}$ and the reduced two-qubit state $\rho^{(5)}_{BC}$ and they are given by
	\begin{eqnarray}
		&&\lambda_{max}(I_{2}\otimes \rho_{BC}^{(5)})=\frac{3+5p_{s}}{12}\nonumber\\&&
		\lambda_{min}(\overline{\rho_{ABC}^{(5)}(I_{2}\otimes \rho_{BC}^{(5)})})=\nonumber\\&&\frac{9+30p_{s}+25p^{2}_{s}-8\sqrt{3}\sqrt{9p_{s}^{2}+14p_{s}^{3}+9p_{s}^{4}}}{288}\nonumber\\&&
		\lambda_{max}(\overline{\rho_{ABC}^{(5)}(I_{2}\otimes \rho_{BC}^{(5)})})=\nonumber\\&&\frac{9+30p_{s}+25p^{2}_{s}+8\sqrt{3}\sqrt{9p_{s}^{2}+14p_{s}^{3}+9p_{s}^{4}}}{288}\nonumber\\&&
		\lambda_{k}(I_{2}\otimes \rho_{BC}^{(5)})=\frac{1-p_{s}}{4}\nonumber\\&&
		Tr[W^{(xy)}_{CHSH}\rho_{BC}^{(5)}(\rho_{BC}^{(5)})^{T_{C}}]=\frac{9-6\sqrt{2}p_{s}+(3-2\sqrt{2})p_{s}^{2}}{18}\nonumber\\&&
		\lambda_{min}[((\rho_{BC}^{(5)})^{T_{C}})^{2}]=  \nonumber\\&& \frac{9-6p_{s}+21p_{s}^{2}-4\sqrt{5}\sqrt{9p_{s}^{2}-6p_{s}^{3}+p_{s}^{4}}}{144} \nonumber\\&& \lambda_{max}((\rho_{BC}^{(5)})^{T_{C}})= \frac{3-p_{s}+2\sqrt{5}p_{s}}{12} \nonumber \\ &&
		\lambda_{max}(W_{CHSH}^{(xy)})=2(1+\sqrt{2}) \nonumber \\ &&
		Tr[((\rho_{BC}^{(5)})^{T_{C}})^{2}]= \frac{9+11p_{s}^{2}}{36}
		\label{info5}
	\end{eqnarray}
	Also, the range of $p$ and $q$ in terms of state parameter $p_{s}$ are given by
	\begin{eqnarray}
		\frac{\sqrt{2}H}{\sqrt{2}H-\frac{3+p_{s}}{12}} < p <  \frac{2H}{2H-\frac{3+p_{s}}{12}}
		\label{ranp5}
	\end{eqnarray}
	and
	\begin{eqnarray}
		\frac{F}{F+2\sqrt{2}\frac{3+p_{s}}{12}} < q <  \frac{F}{F-2\sqrt{2}\frac{3+p_{s}}{12}}
		\label{ranq5}
	\end{eqnarray}
	where $F$ and $H$ given in the previous section can be calculated using the information given in (\ref{info5}).\\
	Therefore, using the information given in (\ref{eigval5}), (\ref{info5}), (\ref{ranp5}), and (\ref{ranq5}), the value of the expression of $S_{v}^{(3)}$ and $S_{v}^{(4)}$ can be calculated for the three-qubit state $\rho_{ABC}^{(5)}$ and they are tabulated in the Table-\ref{t5}.

\section{Comparing our criterion with other existing criteria}
\noindent In this section, we have compared our results with other pre-existing criteria such as (i) M. Li's criterion \cite{mli} (ii) Different types of Svetlichny inequality studied in \cite{bancal}, for the detection of genuine non-locality of pure or mixed three-qubit states. We may re-state M. Li's criterion as \cite{mli}: If $S_{v}$ denote the Svetlichny operator and if any pure or mixed three-qubit states described by the density operator $\varrho$ violate the inequality
\begin{eqnarray}
max|\langle S_{v} \rangle_{\rho}|\leq 4\lambda_{1}
\label{res1} 
\end{eqnarray}
then the state $\rho$ may possess genuine non-local property.\\ Here maximum is taken over all measurement settings and $\lambda_{1}$ denoting the maximum singular value of the matrix $M=[M_{j,ik}]$ with $M_{ijk}=Tr[\rho(\sigma_{i}\otimes \sigma_{j} \otimes \sigma_{k})]$. We can note that in this case, the upper bound given in (\ref{res1}) is state dependent.
\subsection{Example-1} 
\noindent Let us consider a mixed three-qubit state of the form \cite{mli}
\begin{eqnarray}
\rho^{(6)}_{ABC}&=& t |\phi_{gs}\rangle\langle \phi_{gs}|+\frac{1-t}{8}I, ~~ t\in [0,1] 
\label{mstate7}
\end{eqnarray}
where $|\phi_{gs}\rangle=\frac{1}{2}|000\rangle+ \frac{\sqrt{3}}{2}|11\rangle(Cos\theta_{3}|0\rangle +Sin\theta_{3}|1\rangle)$, where $\theta_{3} \in [0,\frac{\pi}{2}]$ and $I_{8 \times 8}$ is an identity matrix of order $8$.\\
The maximum and minimum eigenvalue of $\rho^{(6)}_{ABC}$ is given by
\begin{eqnarray}
\lambda_{max}(\rho_{ABC}^{(6)})=\frac{1+7t}{8},~~\lambda_{min}(\rho_{ABC}^{(6)})=\frac{1-t}{8}
\label{eigval6} 
\end{eqnarray}
It can be observed that if we trace out either system A or system B then the resulting two qubit state will become separable state and thus we cannot apply our result. So, we consider the two-qubit state resulting from tracing out the system C from the three-qubit state $\rho_{ABC}^{(6)}$. The reduced two-qubit state $\rho^{(6)}_{AB}$ is given by
\begin{eqnarray}
\rho^{(6)}_{AB}=\begin{pmatrix} 
	\frac{1}{4} & 0& 0 & \frac{\sqrt{3}t}{4} Cos\theta_{3} \\
			0 & 	\frac{1-t}{4} & 0 & 0\\
			0 & 0 & 	\frac{1-t}{4} & 0\\
			\frac{\sqrt{3}t}{4} Cos\theta_{3} & 0 & 0 & 	\frac{1+2t}{4}\\
		\end{pmatrix}
	\end{eqnarray}
	The state $\rho^{(6)}_{AB}$ is an entangled state for $t\in [0.83,1] ~\text{and}~ \theta_{3} \in [0.615,0.6219]$ as there exists a witness operator $W_{CHSH}^{(xz)}(=2I-B_{CHSH}^{(xz)})$ that detects it. The CHSH witness operator $B_{CHSH}^{(xz)}$ is given by (\ref{bchsh}). This is clear from the following fact 
	\begin{eqnarray}
		Tr[W_{CHSH}^{(xz)}\rho^{(6)}_{AB}]&=&2-\frac{t(2+\sqrt{3}Cos\theta_{3})}{\sqrt{2}}\nonumber\\&<& 0,~\text{for}~t \in [0.83,1] \nonumber\\&& ~\text{\&}~ \theta_{3} \in [0.615,0.6219]
	\end{eqnarray}
	Since the two-qubit state $\rho^{(6)}_{AB}$ is an entangled state and it is detected by $W_{CHSH}^{(xz)}$ so the strength of its non-locality may be measured by $S_{NL}(\rho^{(6)}_{AB})$. It is then given by
	\begin{eqnarray}
		&&S_{NL}(\rho^{(6)}_{AB}) =\frac{-Tr[W_{CHSH}^{(xz)}\rho^{(6)}_{AB}]}{8} \in ~~[0,0.04],\nonumber\\
		&&\text{for} ~t\in [0.83,1]~\text{\&}~ \theta_{3} \in [0.615,0.6219]
		\label{snlinfo6}
	\end{eqnarray}
	Further, we are now in a position to calculate the value of the following expressions involving the three-qubit state $\rho_{ABC}^{(6)}$ and the reduced two-qubit state $\rho^{(6)}_{AB}$. They are given by 
	\begin{eqnarray}
		&&\lambda_{max}(I_{2}\otimes \rho_{AB}^{(6)})=\frac{2+2t+\sqrt{2}t\sqrt{5+3Cos2\theta_{3}}}{8}\nonumber\\&&
		\lambda_{k}(I_{2}\otimes \rho_{AB}^{(6)})=\frac{1-t}{4}
		\label{info6}
	\end{eqnarray}
	Moreover, the range of $p$ and $q$ in terms of state parameter $\theta$ are given by
\begin{eqnarray}
	&&\frac{\sqrt{2}A}{\sqrt{2}A-\frac{2+2t+\sqrt{2}t\sqrt{5+3Cos2\theta_{3}}}{8}} < p \nonumber\\ &<&  \frac{2A}{2A-\frac{2+2t+\sqrt{2}t\sqrt{5+3Cos2\theta_{3}}}{8}}
	\label{ranp6}
\end{eqnarray}
\begin{eqnarray}
\frac{\sqrt{2}B}{\sqrt{2}B+\frac{1-t}{4}}	<  q < \frac{2B}{2B+\frac{1-t}{4}}
	\label{ranq6}
\end{eqnarray}
where $A=(\lambda_{min}(\overline{\rho_{ABC}^{(6)}(I_{2}\otimes \rho_{AB}^{(6)})})+2\lambda_{max}(\rho_{ABC}^{(6)})S_{NL}(\rho_{AB}^{(6)}))$ and $B=(\lambda_{max}(\overline{\rho_{ABC}^{(6)}(I_{2}\otimes \rho_{AB}^{(6)})})+2\lambda_{min}(\rho_{ABC}^{(6)})S_{NL}(\rho_{AB}^{(6)}))$ can be calculated using the information mentioned in (\ref{info6}).\\
	Using the information given in (\ref{eigval6}), (\ref{snlinfo6}), (\ref{info6}), (\ref{ranp6}), and (\ref{ranq6}) the value of the expression of $S_{v}^{(1)}$ and $S_{v}^{(2)}$ can be tabulated for the three-qubit state $\rho_{ABC}^{(6)}$ in Table-\ref{t6}.\\

\noindent We are now in a position to compare our result with the result given in \cite{mli}. We have calculated the maximum singular value  $\lambda_{1}$ of the matrix $M=[M_{j,ik}]$, where $M_{ijk}=Tr[\rho_{ABC}^{(6)}(\sigma_{i}\otimes \sigma_{j} \otimes \sigma_{k})]$ and the values of $\lambda_{1}$ are given in Table-\ref{t6}. It is clear from Table-\ref{t6} that the state $\rho^{(6)}_{ABC}$ with parameters $t\in [0.83,1] ~\text{and}~ \theta_{3} \in [0.615,0.6219]$ violate the bounds given in Result-2 and thus able to detect the genuine non-locality of $\rho^{(6)}_{ABC}$. On the other hand, the state $\rho^{(6)}_{ABC}$ satisfies (\ref{res1}) and thus we can say that M. Li et.al.'s criterion is unable to detect the genuine non-locality of the state $\rho^{(6)}_{ABC}$.
\subsection{Example-2}
\noindent In \cite{bancal}, J.-D. Bancal et. al. have considered a pure state $|psi^{(7)}\rangle_{ABC}$ of the form 
\begin{eqnarray}
|\psi^{(7)}\rangle_{ABC}= \frac{\sqrt{3}}{2}|000\rangle +\frac{\sqrt{3}}{4}|110\rangle + \frac{1}{4} |111\rangle
\label{psi7} 
\end{eqnarray}
The state (\ref{psi7}) is peculiar in the sense that it does not violate 1087 types of Svetlichny Inequality, which have been constructed in \cite{bancal}. Thus, our task is to enquire whether the genuine non-locality of the pure state (\ref{psi7}) is detected by our criterion.\\ 
The pure state (\ref{psi7}) is described by the density operator $\rho_{ABC}^{(7)}= |\psi^{7}\rangle_{ABC}\langle\psi^{7}|$, is an entangled state, and also we have
\begin{eqnarray}
\lambda_{max}(\rho_{ABC}^{(7)})=1,~~\lambda_{min}(\rho_{ABC}^{(7)})=0
\label{eigval7} 
\end{eqnarray}
Tracing out system C from the three-qubit state $\rho_{ABC}^{(7)}$, the reduced two-qubit state $\rho^{(7)}_{AB}$ is given by
	\begin{eqnarray}
		\rho^{(7)}_{AB}=\begin{pmatrix} 
			\frac{3}{4} & 0& 0&\frac{3}{8} \\
			0 & 0 & 0 & 0\\
			0 & 0 & 0 & 0\\
			\frac{3}{8} & 0& 0& \frac{1}{4}\\
		\end{pmatrix}
	\end{eqnarray}
The state $\rho^{(7)}_{AB}$ is an entangled state and it is detected by the witness operator $W_{CHSH}^{(xz)}(=2I-B_{CHSH}^{(xz)})$. It is clear from the following fact
\begin{eqnarray}
Tr[W_{CHSH}^{(xz)}\rho^{(7)}_{AB}]=-0.47487
\end{eqnarray}
The strength of the non-locality of two-qubit state $\rho^{(7)}_{AB}$ may be measured by $S_{NL}(\rho^{(7)}_{AB})$ and it is given by
\begin{eqnarray}
S_{NL}(\rho^{(7)}_{AB}) &=&\frac{-Tr[W_{CHSH}^{(xz)}\rho^{(7)}_{AB}]}{8}\nonumber\\
&=&0.0593588
\label{snlinfo7}
\end{eqnarray}
Further, we can calculate the following information using the three-qubit state $\rho_{ABC}^{(7)}$ and the reduced two-qubit state $\rho^{(7)}_{AB}$ and they are given by
\begin{eqnarray}
&&\lambda_{max}(I_{2}\otimes \rho_{AB}^{(7)})=0.950694\nonumber\\&&
\lambda_{min}(\overline{\rho_{ABC}^{(7)}(I_{2}\otimes \rho_{AB}^{(7)})})=-0.0783743\nonumber\\&&
\lambda_{max}(\overline{\rho_{ABC}^{(7)}(I_{2}\otimes \rho_{AB}^{(7)})})=0.656499\nonumber\\&&
\lambda_{k}(I_{2}\otimes \rho_{AB}^{(7)})=0.0493061
\label{info7}
\end{eqnarray}
Moreover, the range of $p$ and $q$ is given by
\begin{eqnarray}
0.00687286 < p <  0.0096921
\label{ranp7}
\end{eqnarray}
\begin{eqnarray}
0.949571 <  q <  0.963807
\label{ranq7}
\end{eqnarray}
Using the information given in (\ref{eigval7}), (\ref{snlinfo7}), (\ref{info7}), (\ref{ranp7}), and (\ref{ranq7}), the value of the expression of $S_{v}^{(1)}$ and $S_{v}^{(2)}$ can be calculated for the three-qubit state $\rho_{ABC}^{(7)}$ and they are tabulated in Table-\ref{t7}.\\
Therefore, we can infer that for the corresponding $p$ and $q$, the state $|\psi^{(7)}\rangle_{ABC}$ exhibits genuine nonlocality. So, by using our approach, we can say that the state $|\psi^{(7)}\rangle_{ABC}$ may exhibit genuine non-locality.
	
\subsection{Example-3}
\noindent Let us take a mixed three-qubit state of the form \cite{toth}
\begin{eqnarray}
&&\rho^{(8)}_{ABC}= \frac{1}{8} I\otimes I \otimes I+ \sum_{k=x,y,z}\big(\frac{1}{24}(I\otimes\sigma_{k} \otimes \sigma_{k} )\nonumber \\&-&\frac{c}{16}(\sigma_{k}\otimes I \otimes \sigma_{k}+ \sigma_{k}\otimes \sigma_{k} \otimes I )\big),
c \in (0,1] 
\label{s8}
\end{eqnarray}
where $\sigma_{k}$ are the Pauli matrices $k=x,y,z$.
Toth and Acin \cite{toth} have shown that the mixed three-qubit state (\ref{s8}) is a genuine entangled state for $c \in (0.869,1]$ although it admits local hidden variable model. Now we will show that the state $\rho^{(8)}_{ABC}$ violate the bound (\ref{thm5}). To execute this task, let us calculate the maximum and minimum eigenvalue of $\rho_{ABC}^{(8)}$. They are given by
\begin{eqnarray}
\lambda_{max}(\rho_{ABC}^{(8)})=\frac{2+3c}{12}, \lambda_{min}(\rho_{ABC}^{(8)})=0
\label{eigval8} 
\end{eqnarray}
Tracing out system C from the three-qubit state $\rho_{ABC}^{(8)}$, the reduced two-qubit state $\rho^{(8)}_{AB}$ is given by 
\begin{eqnarray}
\rho^{(8)}_{AB}=\begin{pmatrix} 
\frac{2-c}{8} & 0& 0&0 \\
0 & \frac{2+c}{8} & \frac{-c}{4} & 0\\
0 & \frac{-c}{4} & \frac{2+c}{8} & 0\\
0 & 0& 0& \frac{2-c}{8}\\
\end{pmatrix}
\end{eqnarray}
$\rho_{AB}^{(8)}$ is an entangled state for   $c \in (0.869,1]$. Also, we have
\begin{eqnarray}
Tr[W_{CHSH}^{(xy)}\rho^{(8)}_{AB}]=2-\sqrt{2}c ~>0,
~0.869< c \leq 1 \nonumber
\end{eqnarray}
We find that CHSH witness operator $W_{CHSH}^{(xy)}$  is not able to detect the entangled state $\rho^{(8)}_{AB}$.  The strength of the non-locality of $\rho^{(8)}_{AB}$ may be measured by $S_{NL}^{New}(\rho^{(8)}_{AB})$. Therefore, using (\ref{qub1}), the parameter $r$ is given by
\begin{eqnarray}
	r<[0.73,0.815), ~c~\in~(0.869,1]
\end{eqnarray}
Hence, using (\ref{snldef}), the strength of the non-locality $S_{NL}^{New}(\rho^{(8)}_{AB})$ may be calculated as
\begin{eqnarray}
S_{NL}^{New}(\rho^{(8)}_{BC})~~\in~~[0.21,0.44), ~c~\in~(0.869,1]
\end{eqnarray}
Further, we can calculate the following using the three-qubit state $\rho_{ABC}^{(8)}$ and the reduced two-qubit state $\rho^{(8)}_{AB}$
	\begin{eqnarray}
		&&\lambda_{max}(I_{2}\otimes \rho_{AB}^{(8)})=\frac{2+3c}{8}\nonumber\\&&
		\lambda_{min}(\overline{\rho_{ABC}^{(8)}(I_{2}\otimes \rho_{AB}^{(8)})})=0\nonumber\\&&
		\lambda_{max}(\overline{\rho_{ABC}^{(8)}(I_{2}\otimes \rho_{AB}^{(8)})})=-\frac{(c-2)(2+3c)}{96}\nonumber\\&&
		\lambda_{k}(I_{2}\otimes \rho_{AB}^{(8)})=\frac{2-c}{8}\nonumber\\&&
		Tr[W^{(xy)}_{CHSH}\rho_{AB}^{(8)}(\rho_{AB}^{(8)})^{T_{B}}]=\frac{4+2\sqrt{2}c+(1+\sqrt{2})c^{2}}{8}\nonumber\\&&
		\lambda_{min}[((\rho_{AB}^{(8)})^{T_{B}})^{2}]=  \frac{4-12c+9c^{2}}{64} \nonumber\\&& \lambda_{max}((\rho_{AB}^{(8)})^{T_{B}})= \frac{2-c}{8}
		\label{info8}
	\end{eqnarray}
Also, the range of $p$ and $q$ in terms of state parameter $c$ is given by
\begin{eqnarray}
\frac{\sqrt{2}H}{\sqrt{2}H-\frac{2+3c}{8}} < p <  \frac{2H}{2H-\frac{2+3c}{8}}
\label{ranp8}
\end{eqnarray}
\begin{eqnarray}
\frac{F}{F+2\sqrt{2}(\frac{2-c}{8})} <  q <  	\frac{F}{F+2(\frac{2-c}{8})}
\label{ranq8}
\end{eqnarray}
where $H= \frac{(c-2)(2+3c)}{6(4-12c+9c^{2})}$, $F=-\frac{(c-2)(2+3c)}{24}$.\\ 
Using the information given in (\ref{eigval8}), (\ref{info8}), (\ref{ranp8}), and (\ref{ranq8}), the value of the expression of $S_{v}^{(3)}$ and $S_{v}^{(4)}$ can be calculated for the three-qubit state $\rho_{ABC}^{(8)}$ and they are tabulated in the Table-\ref{t8}. 

\section{Conclusion}
	\noindent To summarize, we have considered the problem of detection of non-locality of an arbitrary three-qubit state (pure or mixed). This problem may be handled by the violation of Svetlichny inequality but to do this, we have to maximize the expectation value of the Svetlichny operator overall measurements of unit spin vectors. This optimization problem may not be very simple to solve analytically for an arbitrary three-qubit state. Therefore, we have adopted a new procedure to identify the genuine non-locality of an arbitrary three-qubit state. We have derived a state-dependent lower and upper bound of the expectation value of the Svetlichny operator $S_{v}$ with respect to any pure or mixed three-qubit state described by the density operator $\rho_{ABC}$. These bounds established a connection between $\langle S_{v} \rangle_{\rho_{ABC}}$ and the strength of the non-locality of the reduced two-qubit entangled state $\rho_{ij},~i\neq j,~i,j=A,B,C$. We should note here that the considered reduced two-qubit state must be an entangled state. The strength of the non-locality of the reduced two-qubit state may be measured either by $S_{NL}(\rho_{ij})$ or by $S^{New}_{NL}(\rho_{ij})$ depending on whether it is detected or not detected by CHSH witness operator. We have shown that the obtained lower and upper bound of the expectation value of the Svetlichny operator may help in deriving the inequality violation which shows the genuine non-locality of the three-qubit state under investigation. To implement our results in an experiment, let us discuss briefly the possible implementation of the partial trace, eigenvalues, and partial transposition in an experiment: (i) Partial Trace-  It is a very common function of the composite system. It is not only viewed as a  mathematical operation but  also has operational meaning \cite{paris,garola,fortin}. The numerical calculation of the partial trace function has been presented and has shown that it may be implemented using Bloch's parametrization with generalized Gell Mann's matrices \cite{maziero}. (ii) Eigenvalues- It is shown that there exist methods by which one may determine the eigenvalues of a state experimentally in a relatively easier way than full state tomography \cite{ekert,tanaka}. (iii) Partial Transposition-  Partially transposed density matrices are generically unphysical because it is a positive map but not completely positive but in spite of this limitation, measurement of their moments is possible \cite{gray}. Using their moments, one may estimate the values of the trace of a function of partial transposition \cite{huang}.
	
	\section{Acknowledgement}
	\noindent A. G. would like to acknowledge the financial support from CSIR. This work is supported by CSIR File No. 08/133(0035)/2019-EMR-1.

	\section{DATA AVAILABILITY STATEMENT}
\noindent Data sharing not applicable to this article as no datasets were generated or analysed during the current study.
	
	\section{Appendix}
	\subsection{Appendix-I}
	\noindent \textbf{Proof of Lemma-1:} Let us start with the quantity  $R(Tr[W_{CHSH}\rho_{ij}(\rho_{ij}^{T_{j}})^{2}])$. Applying LHS of \textbf{Cor-1} on Hermitian operator $(\rho_{ij}^{T_{j}})^{2}$ and $W_{CHSH}\rho_{ij}$ be any complex matrix, we get
	\begin{eqnarray}
		R(Tr[W_{CHSH}\rho_{ij}(\rho_{ij}^{T_{j}})^{2}])&\geq& \lambda_{min}[(\rho_{ij}^{T_{j}})^{2}]\times \nonumber\\&& Tr[\overline{W_{CHSH}\rho_{ij}}]
		\label{l1}
	\end{eqnarray}
where, $Tr[\overline{W_{CHSH}\rho_{ij}}]=Tr[W_{CHSH}\rho_{ij}]$.\\	Again applying RHS of \textbf{Cor-1} on Hermitian operator $\rho_{ij}^{T_{j}}$ and $W_{CHSH}\rho_{ij}\rho_{ij}^{T_{j}}$ be any complex matrix, the quantity $R(Tr[W_{CHSH}\rho_{ij}(\rho_{ij}^{T_{j}})^{2}])$ can also be expressed as 
	\begin{eqnarray}
		R(Tr[W_{CHSH}\rho_{ij}(\rho_{ij}^{T_{j}})^{2}])&\leq& \lambda_{max}[\rho_{ij}^{T_{j}}]Tr[\overline{W_{CHSH}\rho_{ij}\rho_{ij}^{T_{j}}}]\nonumber\\
		&=& 4\lambda_{max}[\rho_{ij}^{T_{j}}]N(\rho_{ij})K	
		\label{lemma1}
	\end{eqnarray}
	Since, $Tr[\overline{W_{CHSH}\rho_{ij}\rho_{ij}^{T_{j}}}]$ = $Tr[W_{CHSH}\rho_{ij}\rho_{ij}^{T_{j}}]$. So, in the second line of $(\ref{lemma1})$, we have used the relation $(\ref{k1})$ i.e. $Tr[W_{CHSH}\rho_{ij}\rho_{ij}^{T_{j}}]=4N(\rho_{ij})K$.\\
	Using (\ref{l1}), the equation (\ref{lemma1}) can be re-expressed as 
	\begin{eqnarray}
		K\geq\frac{\lambda_{min}[(\rho_{ij}^{T_{j}})^{2}]Tr[W_{CHSH}\rho_{ij}]}{4\lambda_{max}[\rho_{ij}^{T_{j}}]N(\rho_{ij})}
	\end{eqnarray}
	Hence proved.\\
	
	\subsection{Appendix-II}
	\noindent \textbf{Lemma-2:} If $\rho_{ABC}$ denote an arbitrary three-qubit state and $\rho_{ij},i,j=A,B,C,i\neq j$ be its reduced two-qubit entangled state, which is not detected by CHSH witness operator 
	$W_{CHSH}$ then the non-locality of $\rho_{ij}$ may be determined by $S^{New}_{NL}(\rho_{ij})$ given in (\ref{snldef}). The quantity $K$ involved in the expression of $S^{New}_{NL}(\rho_{ij})$ is bounded above and its upper bound is given by 	
	\begin{eqnarray}
		K\leq \frac{\lambda_{max}(W_{CHSH})\lambda_{max}(\rho_{ij})Tr[W_{CHSH}\rho_{ij}]+Tr[(\rho_{ij}^{T_{j}})^{2}]}{8N(\rho_{ij})} 
		\label{K1}
	\end{eqnarray}
	\textbf{Proof:}	Let us consider the two operators given by
	\begin{eqnarray}
		A_{2}=W_{CHSH}\rho_{ij},~~B_{2}=\rho_{ij}^{T_{j}}
		\label{a2b2}
	\end{eqnarray}
	For the two operators $A_{2}$ and $B_{2}$ defined in (\ref{a2b2}), we have
	\begin{eqnarray}
		&&(A_{2}-B_{2})^{2}\geq 0 \nonumber\\&&
		\implies A_{2}^{2}-A_{2}B_{2}-B_{2}A_{2}+B_{2}^{2}\geq0
		\label{ineq110} 
	\end{eqnarray}
	Taking trace both sides of (\ref{ineq110}) and simplifying, we get
	\begin{eqnarray}
		2Tr(A_{2}B_{2})\leq Tr(A_{2}^{2})+Tr(B_{2}^{2})
		\label{ineq111} 
	\end{eqnarray}
	Using (\ref{a2b2}) and (\ref{ineq111}), we get
	\begin{eqnarray}
		2Tr(W_{CHSH}\rho_{ij}\rho_{ij}^{T_{j}})&\leq& Tr((W_{CHSH}\rho_{ij})^{2})\nonumber\\&+&Tr((\rho_{ij}^{T_{j}})^{2})
		\label{ineq112} 
	\end{eqnarray}
	Also, applying \textbf{Cor-1} on Hermitian operator  $W_{CHSH}$ and considering $W_{CHSH}(\rho_{ij})^{2}$ be any complex matrix, and using the fact that $Tr[W_{CHSH}(\rho_{ij})^{2}]= Tr[\overline{W_{CHSH}(\rho_{ij})^{2}}]$ we get
	\begin{eqnarray}
		Tr((W_{CHSH}\rho_{ij})^{2}) &\leq& \lambda_{max}(W_{CHSH})\times \nonumber\\&& Tr[W_{CHSH}(\rho_{ij})^{2}]
		\label{ineq113} 
	\end{eqnarray}
Again applying \textbf{Cor-1} on Hermitian operator  $\rho_{ij}$ and $W_{CHSH}\rho_{ij}$ be any complex matrix, and using the fact that $Tr[W_{CHSH}\rho_{ij}]= Tr[\overline{W_{CHSH}\rho_{ij}}]$
\begin{eqnarray}
Tr[W_{CHSH}(\rho_{ij})^{2}] &\leq& \lambda_{max}(\rho_{ij}) Tr[W_{CHSH}\rho_{ij}]
	\label{ineqnew} 
\end{eqnarray}
Using (\ref{ineq113}) and (\ref{ineqnew}), we get 
\begin{eqnarray}
	Tr((W_{CHSH}\rho_{ij})^{2}) &\leq& \lambda_{max}(W_{CHSH})\times \nonumber\\&&\lambda_{max}(\rho_{ij}) Tr[W_{CHSH}\rho_{ij}]
	\label{ineqnew1} 
\end{eqnarray}
	Using (\ref{ineq112}) and (\ref{ineqnew1}), we get
	\begin{eqnarray}
		~~~2Tr(W_{CHSH}\rho_{ij}\rho_{ij}^{T_{j}})&\leq& \lambda_{max}(W_{CHSH})\lambda_{max}(\rho_{ij})\times \nonumber\\&&Tr[W_{CHSH}\rho_{ij}] +Tr[(\rho_{ij}^{T_{j}})^{2}] \nonumber\\
		\label{ineq114}
	\end{eqnarray}
	Putting $Tr[W_{CHSH}\rho_{ij}(\rho_{ij})^{T_{j}}]=4N(\rho_{ij})K$ in (\ref{ineq114}), we get 
	\begin{eqnarray}
		K\leq \frac{\lambda_{max}(W_{CHSH})\lambda_{max}(\rho_{ij})Tr[W_{CHSH}\rho_{ij}]+Tr[(\rho_{ij}^{T_{j}})^{2}]}{8N(\rho_{ij})} \nonumber
	\end{eqnarray}

	\subsection{Appendix-III}
	\noindent \textbf{Proof of Theorem-2b:} Let us consider a three-qubit state $\rho_{ABC}$ which satisfies the Svetlichny inequality given by (\ref{svin100}). Now, if a three-qubit state $\rho_{ABC}$ satisfies the Svetlichny inequality then our task is to construct the operator $A_{u}$. To accomplish this task, we need to specify the parameter $q$. Thus, recalling the upper bound of the expectation value of the Svetlichny operator $S_{v}$ given in (\ref{ubs}) and using (\ref{svin100}), the restriction on $q$ may be obtained by solving the inequality
	\begin{eqnarray}
		-4 &\leq&  \frac{8(1-q)\lambda_{max}(\overline{\rho_{ABC}(I_{2}\otimes \rho_{ij})})}{q\lambda_{k}(I_{2}\otimes \rho_{ij})}\nonumber \\ 
		&+&\frac{16(1-q)S_{NL}(\rho_{ij})\lambda_{min}(\rho_{ABC})}{q\lambda_{k}(I_{2}\otimes \rho_{ij})}\leq 4
		\label{ineq2} 
	\end{eqnarray} 
	Solving the L.H.S. of inequality (\ref{ineq2}) for the parameter $q$ and simplifying, we get 
	\begin{eqnarray}
		q\geq l_{2}=\frac{2}{d^{(+)}_{2}}&\times& [\lambda_{max}(\overline{\rho_{ABC}(I_{2}\otimes \rho_{ij})}) + 2S_{NL}(\rho_{ij})\nonumber\\&&\lambda_{min}(\rho_{ABC})] \nonumber 
	\end{eqnarray}
	where $d^{(+)}_{2}=2[\lambda_{max}(\overline{\rho_{ABC}(I_{2}\otimes \rho_{ij})})+ 2S_{NL}(\rho_{ij})\lambda_{min}(\rho_{ABC})]+\lambda_{k}(I_{2}\otimes\rho_{ij})$.\\
	Then, by solving the L.H.S. of the inequality (\ref{ineq2}), we get $q\geq 1$ which is not possible. Thus, considering $q \leq min\{\frac{2}{d^{(-)}_{2}}\times [\lambda_{max}(\overline{\rho_{ABC}(I_{2}\otimes \rho_{ij})}) + 2S_{NL}(\rho_{ij})\lambda_{min}(\rho_{ABC})],1\}$, where $d^{(-)}_{2}=2[\lambda_{max}(\overline{\rho_{ABC}(I_{2}\otimes \rho_{ij})})+ 2S_{NL}(\rho_{ij})\lambda_{min}(\rho_{ABC})]-\lambda_{k}(I_{2}\otimes\rho_{ij})$, we get the required result. Hence proved.
	\subsection{Appendix-IV}
	\noindent \textbf{Proof of Theorem-3a:} Let us consider a three-qubit state $\rho_{ABC}$ which satisfies the Svetlichny inequality (\ref{svin100}). Now, if a three-qubit state $\rho_{ABC}$ satisfies the Svetlichny inequality then our task is to construct the operator $A_{l}$. To accomplish this task, we need to specify the parameter $p$. Thus, recalling the lower bound of the expectation value of the Svetlichny operator $S_{v}$ given in (\ref{svlb2}) and using (\ref{svin100}), the restriction on $p$ may be obtained by solving the inequality
	\begin{eqnarray}
		-4 &\leq& 8(1-p)\biggl[\frac{\lambda_{min}(\overline{\rho_{ABC}(I_{2}\otimes \rho_{ij})})}{p\lambda_{max}(I_{2}\otimes \rho_{ij})}-\nonumber\\&&\frac{\big(S^{New}_{NL}(\rho_{ij})-r(P^{max}-\frac{3}{4})\times A_{1}\big)}{p(1-r)\lambda_{min}[(\rho_{ij}^{T_{j}})^{2}]\lambda_{max}(I_{2}\otimes \rho_{ij})}\biggr]\nonumber \\&\leq& 4
		\label{ineq3} 
	\end{eqnarray} 
where $A_{1}=(N(\rho_{ij})\lambda_{max}(\rho_{ij}^{T_{j}})\lambda_{max}(\rho_{ABC}))$.\\
	Solving the L.H.S. of inequality (\ref{ineq3}) for the parameter $p$ and simplifying, we get 
	\begin{eqnarray}
		p\geq l_{3}= \frac{2H}{2H-\lambda_{max}(I_{2}\otimes \rho_{ij})}
		\label{l3}
	\end{eqnarray}
	where $H= \lambda_{min}(\overline{\rho_{ABC}(I_{2}\otimes \rho_{ij})})-(\frac{S^{New}_{NL}(\rho_{ij})-r(P^{max}-\frac{3}{4})}{(1-r)\lambda_{min}[(\rho_{ij}^{T_{j}})^{2}]})\times (N(\rho_{ij})\lambda_{max}(\rho_{ij}^{T_{j}})\lambda_{max}(\rho_{ABC}))$.\\
	Then, by solving the R.H.S. of the inequality (\ref{ineq3}), we get $p\geq 1$ which is not possible. Thus, considering $p \leq min\{\frac{2H}{2H-\lambda_{max}(I_{2}\otimes \rho_{ij})},1\}$, we get the required result. Hence proved.\\	
	
	\subsection{Appendix-V}
	\noindent \textbf{Proof of Theorem-3b:} Let us consider a three-qubit state $\rho_{ABC}$ which satisfies the Svetlichny inequality (\ref{svin100}). Now, if a three-qubit state $\rho_{ABC}$ satisfies the Svetlichny inequality then our task is to construct the operator $A_{u}$. To accomplish this task, we need to specify the parameter $q$. Thus, recalling the upper bound of the expectation value of the Svetlichny operator $S_{v}$ given in (\ref{ubs1}) and using (\ref{svin100}), the restriction on $q$ may be obtained by solving the inequality
	\begin{eqnarray}
		-4 &\leq&  \frac{2(1-q)}{q\lambda_{k}(I_{2}\otimes \rho_{ij})}
		\biggl[4\lambda_{max}(\overline{\rho_{ABC}(I_{2}\otimes \rho_{ij})})\nonumber \\&-&\frac{\lambda_{min}(\rho_{ABC})\times A_{2}}{\lambda_{max}(W_{CHSH})\lambda_{max}(\rho_{ij})}\biggr] \leq 4
		\label{ineq4} 
	\end{eqnarray} 
where $A_{2}= \frac{8N(\rho_{ij})(S^{New}_{NL}(\rho_{ij})-r(P^{max}-\frac{3}{4}))}{1-r}-Tr[(\rho_{ij}^{T_{j}})^{2}]$.\\
	Solving the R.H.S. of inequality (\ref{ineq4}) for the parameter $q$ and simplifying, we get 
	\begin{eqnarray}
		q \geq l_{4}=\frac{F}{F+ 2\lambda_{k}(I_{2}\otimes \rho_{ij})}
		\label{l41}
	\end{eqnarray}
	where $F=(4\lambda_{max}(\overline{\rho_{ABC}(I_{2}\otimes\rho_{ij})})-\frac{\lambda_{min}(\rho_{ABC})}{\lambda_{max}(W_{CHSH})\lambda_{max}(\rho_{ij})}(\frac{8N(\rho_{ij})(S^{New}_{NL}(\rho_{ij})-r(P^{max}-\frac{3}{4}))}{1-r}-Tr[(\rho_{ij}^{T_{j}})^{2}]))$.\\
	Then, by solving the L.H.S. of the inequality (\ref{ineq4}), we get $q\geq 1$ which is not possible. Thus, considering $q \leq min\{\frac{F}{F- 2\lambda_{k}(I_{2}\otimes \rho_{ij})},1\}$, we get the required result. Hence proved.\\

\begin{table*}[!htbp]
	\begin{tabular} {|c|c|c|c|c|}\hline 
		State parameter & Operator parameter & Operator parameter &  $\langle S_{v}^{(1)}\rangle_{\rho^{(1)}_{ABC}}$ & $\langle S_{v}^{(2)}\rangle_{\rho^{(1)}_{ABC}}$ \\ 
		$(\lambda_{0})$ & $(p)$ & $(q)$ &   &  \\ \hline
		
		0.92 & 0.05 & 0.95 & -4.98604 & 5.4752 \\ \hline 
		0.93 & 0.019 & 0.97 & -5.2497 & 4.70144 \\ \hline 
		0.94 & 0.012 & 0.98 & 4.62363 & 5.48997 \\ \hline 
		0.95 & 0.04 & 0.9943 & 5.06912 & 5.62141 \\ \hline  				
	\end{tabular}
	\caption{We have chosen different values of the three-qubit state parameter $\lambda_{0}$ for which its reduced two-qubit state is entangled. Then we get a value of $S_{NL}(\rho^{(1)}_{BC})$ and corresponding to it, we have chosen a value of the parameters $p$ and $q$ given in (\ref{ranp1}) and (\ref{ranq1}) respectively. Using the information given in (\ref{info1}) and considering few values of $p$, $q$ and $\lambda_{0}$, Table-I is prepared. It depicts the values of $\langle S_{v}^{(1)}\rangle_{\rho_{ABC}^{(1)}}$ \& $\langle S_{v}^{(2)}\rangle_{\rho_{ABC}^{(1)}}$ given in (\ref{slb1}) and (\ref{sub1}) indicating the fact that the state $|\psi^{(1)}\rangle_{ABC}$ exhibit genuine nonlocality.}
	\label{t1}
\end{table*}

\begin{table*}[!htbp]
	\begin{tabular} {|c|c|c|c|c|}\hline 
		State Parameter & Operator parameter & Operator parameter &  $\langle S_{v}^{(1)}\rangle_{\rho^{(2)}_{ABC}}$ & $\langle S_{v}^{(2)}\rangle_{\rho^{(2)}_{ABC}}$ \\ 
		$(t)$ & $(p)$ & $(q)$ &   &  \\ \hline 
		0.55 & 0.006 & 0.48 & -4.96125 & 5.40786 \\ \hline 
		0.65 & 0.019 & 0.44 & -4.5171 & 4.75185 \\ \hline 
		0.72 & 0.025 & 0.38 & -5.32538 & 5.04819 \\ \hline 
		0.79 & 0.039 & 0.46 & -4.77032 & 4.54815 \\ \hline  				
	\end{tabular}
	\caption{We have chosen different values of the three-qubit state parameter $t$ for which its reduced two-qubit state is entangled. Then we get a value of $S_{NL}(\rho^{(2)}_{BC})$ and corresponding to it, we have chosen a value of the operator parameters $p$ and $q$ given in (\ref{ranp2}) and (\ref{ranq2}) respectively. Using the information given in (\ref{info2}) and considering few values of $p$, $q$ and $t$, Table-II is prepared. It depicts the values of $\langle S_{v}^{(1)}\rangle_{\rho_{ABC}^{(2)}}$ \& $\langle S_{v}^{(2)}\rangle_{\rho_{ABC}^{(2)}}$ given in (\ref{slb1}) and (\ref{sub1}) indicating the fact that the state $\rho^{(2)}_{ABC}$ exhibit genuine nonlocality.}
	\label{t2}
\end{table*}

\begin{table*}[!htbp]
	\begin{tabular} {|c|c|c|c|c|}\hline 
		State Parameter & Operator parameter & Operator parameter &  $\langle S_{v}^{(3)}\rangle_{\rho^{(3)}_{ABC}}$ & $\langle S_{v}^{(4)}\rangle_{\rho^{(3)}_{ABC}}$ \\ 
		$(\theta)$ & $(p)$ & $(q)$ &   & \\ \hline
		1.2 & 0.67 & 0.86 & -4.64621 & 5.03388 \\ \hline 
		1.3 & 0.78 & 0.79 & -5.24561 & 4.75133 \\ \hline 
		1.4 & 0.915 & 0.69 & -4.81717 & 5.53806 \\ \hline 
		1.5 & 0.985 & 0.64 & -5.333 & 5.25777 \\ \hline  				
	\end{tabular}
	\caption{We have chosen different values of the three-qubit state parameter $\theta$ for which its reduced two-qubit state is entangled. Then we get a value of $S^{New}_{NL}(\rho^{(3)}_{BC})$ and corresponding to it, we have chosen a value of the operator parameters $p$ and $q$ given in (\ref{ranp3}) and (\ref{ranq3}) respectively. Using the information given in (\ref{info3}) and considering a few values of $p$, $q$, and $\theta$, Table-III is prepared. It depicts the values of $\langle S_{v}^{(3)}\rangle_{\rho_{ABC}^{(3)}}$ \& $\langle S_{v}^{(4)}\rangle_{\rho_{ABC}^{(3)}}$ given in (\ref{slb3}) and (\ref{sub4}) indicating the fact that the state $\rho^{(3)}_{ABC}$ exhibit genuine nonlocality.}
	\label{t3}
\end{table*}

\begin{table*}[!htbp]
	\begin{tabular} {|c|c|c|c|c|}\hline 
		State Parameter & Operator parameter & Operator parameter &  $\langle S_{v}^{(3)}\rangle_{\rho^{(4)}_{ABC}}$ & $\langle S_{v}^{(4)}\rangle_{\rho^{(4)}_{ABC}}$ \\ 
		$(p_{s})$ & $(p)$ & $(q)$ &   &  \\ \hline
		0.5 & 0.9 & 0.75 & -5.3768 & 4.99623 \\ \hline 
		0.6 & 0.95 & 0.8 & -5.46426 & 4.39912 \\ \hline 
		0.7 & 0.98 & 0.82 & -5.03592 & 4.96321 \\ \hline 
		0.8 & 0.993 & 0.86 & -5.11832 & 5.48576 \\ \hline  				
	\end{tabular}
	\caption{We have chosen different values of the three-qubit state parameter $p_{s}$ for which its reduced two-qubit state is entangled. Then we get a value of $S^{New}_{NL}(\rho^{(4)}_{BC})$ and corresponding to it, we have chosen a value of the operator parameters $p$ and $q$ given in (\ref{ranp4}) and (\ref{ranq4}) respectively. Using the information given in (\ref{info4}) and considering a few values of $p$, $q$, and $p_{s}$, Table-IV is prepared. It depicts the values of $\langle S_{v}^{(3)}\rangle_{\rho^{(4)}_{ABC}}$ \& $\langle S_{v}^{(4)}\rangle_{\rho^{(4)}_{ABC}}$ given in (\ref{slb3}) and (\ref{sub4}) indicating the fact that the state $\rho^{(4)}_{ABC}$ exhibit genuine nonlocality.}
	\label{t4}
\end{table*}

\begin{table*}[!htbp]
	\begin{tabular} {|c|c|c|c|c|}\hline 
		State Parameter & Operator parameter & Operator parameter &  $\langle S_{v}^{(3)}\rangle_{\rho^{(5)}_{ABC}}$ & $\langle S_{v}^{(4)}\rangle_{\rho^{(5)}_{ABC}}$ \\ 
		$(p_{s})$ & $(p)$ & $(q)$ &   &  \\ \hline
		0.82 & 0.72 & 0.93 & -4.35959 & 5.06538 \\ \hline 
		0.87 & 0.6 & 0.95 & -4.6602 & 5.30281 \\ \hline 
		0.92 & 0.45 & 0.97 & -5.4101 & 5.45763 \\ \hline 
		0.97 & 0.35 & 0.99 & -5.147 & 5.10813 \\ \hline  				
	\end{tabular}
	\caption{We have chosen different values of the three-qubit state parameter $p_{s}$ for which its reduced two-qubit state is entangled. Then we get a value of $S^{New}_{NL}(\rho^{(5)}_{BC})$ and corresponding to it, we have chosen a value of the operator parameters $p$ and $q$ given in (\ref{ranp5}) and (\ref{ranq5}) respectively. Using the information given in (\ref{info5}) and considering a few values of $p$, $q$, and $p_{s}$, Table-V is prepared. It depicts the values of $\langle S_{v}^{(3)}\rangle_{\rho^{(5)}_{ABC}}$ \& $\langle S_{v}^{(4)}\rangle_{\rho^{(5)}_{ABC}}$ given in (\ref{slb3}) and (\ref{sub4}) indicating the fact that the state $\rho^{(5)}_{ABC}$ exhibit genuine nonlocality.}
		\label{t5}
\end{table*}

\begin{table*}[!htbp]
	\begin{tabular} {|c|c|c|c|c|c|c|}\hline 
		\multicolumn {7}{|c|} {Comparison Analysis}\\
		\hline
		&	\multicolumn {4}{|c|}{Our Work} & \multicolumn {2}{|c|}{ M.Li et.al. Work\cite{mli} }\\
		\hline
		State Parameter & Operator parameter &  $\langle S_{v}^{(1)}\rangle_{\rho^{(6)}_{ABC}}$ & $\langle S_{v}^{(2)}\rangle_{\rho^{(6)}_{ABC}}$ & Whether SI satisfied & Upper Bound of $\langle S_{v} \rangle_{\rho^{(6)}_{ABC}}$ &  Whether SI satisfied  \\ 
		$t,\theta_{3}$ & $p,q$   &  & & or violated?&$= 4\lambda_{1}$ & or violated? \\ \hline 
		$t=0.84$,$\theta_{3}=0.616$ & $p=0.1$,$q=0.92$ & -5.07155 & 5.07541 & Violated& 3.36062 & Satisfied \\ \hline 
		$t=0.87$,$\theta_{3}=0.618$ & $p=0.09$,$q=0.93$ & -4.76827 & 5.65134  & Violated&3.4831 & Satisfied\\ \hline 
		$t=0.9$,$\theta_{3}=0.62$ & $p=0.07$,$q=0.959$ & -5.02466 & 4.35569 &Violated& 3.60576 & Satisfied\\ \hline 
		$t=0.95$,$\theta_{3}=0.6205$ & $p=0.04$,$q=0.979$ & -5.19447 & 4.66488 &Violated& 3.80675 & Satisfied\\ \hline
		$t=0.99$,$\theta_{3}=0.6215$ & $p=0.019$,$q=0.996$ & -4.403 & 4.59326  &Violated& 3.96844 & Satisfied\\ \hline 
		$t=0.998$,$\theta_{3}=0.6216$ & $p=0.012$,$q=0.9992$ & -4.83292 & 4.62339  &Violated& 4.00064 & May violate\\ \hline 
		$t=0.999$,$\theta_{3}=0.6217$ & $p=0.01$,$q=0.9996$ & -5.4911 & 4.62769  &Violated& 4.0048 & May violate\\ \hline   				
	\end{tabular}
	\caption{We have chosen different values of the three-qubit state parameter $(t,\theta_{3})$ for which its reduced two-qubit state $\rho^{(6)}_{AB}$ is entangled. Then we get a value of $S_{NL}(\rho^{(6)}_{AB})$ and corresponding to it, we have chosen a value of the operator parameters $p$ and $q$ given in (\ref{ranp6}) and (\ref{ranq6}) respectively. . Using the information given in (\ref{info6}) and considering few values of $p$, $q$, $t$  and $\theta_{3}$, Table-VI is prepared. It depicts the values of $\langle S_{v}^{(1)}\rangle_{\rho_{ABC}^{(6)}}$ \& $\langle S_{v}^{(2)}\rangle_{\rho_{ABC}^{(6)}}$ given in (\ref{slb1}) and (\ref{sub1}) indicating the fact that the state $\rho^{(6)}_{ABC}$ exhibit genuine nonlocality.}
		\label{t6}
\end{table*}

\begin{table*}[!htbp]
	\begin{tabular} {|c|c|c|c|}\hline 
		Operator parameter & Operator parameter &  $\langle S_{v}^{(1)}\rangle_{\rho^{(7)}_{ABC}}$ & $\langle S_{v}^{(2)}\rangle_{\rho^{(7)}_{ABC}}$ \\ 
		$(p)$ & $(q)$ &   &  \\ \hline
		0.007 & 0.95 & -5.5534 & 5.60622 \\ \hline 
		0.0075 & 0.955 & -5.18056 & 5.01918 \\ \hline 
		0.008 & 0.959 & -4.85433 & 4.55396 \\ \hline 
		0.0085 & 0.96 & -4.56648 & 4.43826 \\ \hline  
			0.009 & 0.963 & -4.31061 & 4.0926 \\ \hline				
	\end{tabular}
	\caption{We traced out system C from $\rho_{ABC}$ and got a reduced two-qubit state which is entangled. Then we get a value of $S_{NL}(\rho^{(7)}_{AB})$ and corresponding to it, we have chosen a value of the parameters $p$ and $q$ given in (\ref{ranp7}) and (\ref{ranq7}) respectively. Using the information given in (\ref{info7}) and considering few values of $p$, and $q$, Table-VII is prepared. It depicts the values of $\langle S_{v}^{(1)}\rangle_{\rho_{ABC}^{(7)}}$ \& $\langle S_{v}^{(2)}\rangle_{\rho_{ABC}^{(7)}}$ given in (\ref{slb1}) and (\ref{sub1}) indicating the fact that the state $|\psi^{(7)}\rangle_{ABC}$ exhibit genuine nonlocality.}
		\label{t7}
\end{table*}

\begin{table*}[!htbp]
	\begin{tabular} {|c|c|c|c|c|}\hline 
		State Parameter & Operator parameter & Operator parameter &  $\langle S_{v}^{(3)}\rangle_{\rho^{(8)}_{ABC}}$ & $\langle S_{v}^{(4)}\rangle_{\rho^{(8)}_{ABC}}$ \\ 
		$(c)$ & $(p)$ & $(q)$ &   &  \\ \hline
		0.87 & 0.89 & 0.36 & -4.14775 & 5.4637 \\ \hline 
		0.89 & 0.87 & 0.37 & -4.15282 & 5.30108 \\ \hline 
		0.92 & 0.82 & 0.38 & -4.73136 & 5.17754 \\ \hline 
		0.95 & 0.8 & 0.43 & -4.29487 & 4.28605 \\ \hline 
		0.99 & 0.76 & 0.44 & -4.14294 & 4.21697 \\ \hline 				
	\end{tabular}
	\caption{We have chosen different values of the three-qubit state parameter $c$ for which its reduced two-qubit state is entangled. Then we get a value of $S^{New}_{NL}(\rho^{(8)}_{BC})$ and corresponding to it, we have chosen a value of the operator parameters $p$ and $q$ given in (\ref{ranp8}) and (\ref{ranq8}) respectively. Using the information given in (\ref{info8}) and considering few values of $p$, $q$ and $c$, Table-VIII is prepared. It depicts the values of $\langle S_{v}^{(3)}\rangle_{\rho^{(8)}_{ABC}}$ \& $\langle S_{v}^{(4)}\rangle_{\rho^{(8)}_{ABC}}$ given in (\ref{slb3}) and (\ref{sub4}) indicating the fact that the state $\rho^{(8)}_{ABC}$ exhibit genuine nonlocality.}
		\label{t8}
\end{table*}

\end{document}